\documentclass[12pt]{article}
\usepackage{amsmath}
\usepackage{tensor}
\usepackage{amssymb}
\usepackage{mathrsfs}
\usepackage{tikz}
\usepackage{hyperref} 
\usepackage{slashed}
\usepackage{enumerate}
\usepackage{cite}
\numberwithin{equation}{section}
\title{Gravitational analog of the Non--Abelian T--duality}
\author{Oleg Lunin\footnote{olunin@albany.edu}\ \ and Parita Shah\footnote{pshah3@albany.edu}}
\date{}
\begin{document}
\def\be{\begin{equation}}
\def\bea{\begin{eqnarray}}
\def\ee{\end{equation}}
\def\eea{\end{eqnarray}}
\def\d{\partial}
\def\eps{\varepsilon}
\def\la{\lambda}
\def\b{\bigskip}
\def\nn{\nonumber \\}
\def\p{\partial}
\def\t{\tilde}
\def\h{{1\over 2}}
\def\be{\begin{equation}}
\def\bea{\begin{eqnarray}}
\def\ee{\end{equation}}
\def\eea{\end{eqnarray}}
\def\b{\bigskip}
\def\u{\uparrow}
\newcommand{\comment}[2]{#2}
\maketitle

\begin{center}
\ \vskip -1.2cm
{\em  Department of Physics,\\
University at Albany (SUNY),\\
1400 Washington Avenue,\\
Albany, NY 12222, USA
 }
 \end{center}

\vskip 0.08cm
\begin{abstract}
Non--abelian T duality (NATD) is a symmetry of the worldsheet action that allows one to generate new solutions of string theory by performing algebraic transformations of known geometries. Applications of such transformations to spheres have been especially fruitful in the past, but the results did not reduce to the abelian transformation in the decompactification limit unless the dimension of sphere was equal to three. We propose a unique counterpart of NATD in classical gravity that reproduces the correct decompactification limit for all spheres $S^n$ but generates an $n$--form field strength and therefore cannot be naturally embedded in a worldsheet theory of NS--NS fields unless $n=3$. We also propose a non--abelian version of the TsT transformation which produces solutions of type II supergravity describing continuous deformations of geometries with $S^n\times S^n$ and $S^n\times T^n$ factors.

\end{abstract}
\newpage
\tableofcontents

\newpage

\section{Introduction}

Over the last three decades, studies of dualities in string theory have led to many remarkable results, ranging from discovery of D--branes \cite{PolchW1,PolchW2,PolchW3,PolchW4,PolchW5} and M--theory \cite{MtheoryW1,MtheoryW2,MtheoryW3} to insights into physics of black holes \cite{BHtdualW1,BHtdualW2,BHtdualW3,BHtdualW4} and AdS/CFT correspondence \cite{AdSCFTw1,AdSCFTw2,AdSCFTw3,AdSCFTw4}.
A prominent role in this progress has been played by T dualities, which establish equivalence between string theories on very different backgrounds by performing nonlocal field transformations on the string worldsheet. This procedure relies on symmetries of the original theory, which can be either abelian or non--abelian. In the former case, the duality can be summarized in a set of universal Buscher rules \cite{Buscher}, while the non--abelian T duality (NATD) has been carried out only on the case--by--case basis \cite{NATD80w1,NATD80w2,OldNATDw1,OldNATDw2,OldNATDw3,OldNATDw4,
NTDgrp,NATD101,NATDAdSCFTw1,NATDAdSCFTw2,NATDAdSCFTw3,
NATDAdSCFTw4,NATDAdSCFTw5,NATDAdSCFTw6,NATDAdSCFTw7,NATDAdSCFTw8,
NATDAdSCFTw9,NATDAdSCFTw10,NATDAdSCFTw11,NATDAdSCFTw12}. These efforts have focused on two distinct situations where the space is a manifold describing either a Lie group $G$ \cite{NTDgrp} or a coset $G/H$ \cite{NTDcosetW1,NTDcosetW2,NTDcosetW3}. In both cases, the duality is performed along the group $G$, so it is only the group prescription that reduces to the standard T duality in the abelian limit\footnote{The details of this limit will be discussed later, but one can think of it as focusing on a vicinity of a particular point on a manifold and pushing the rest of the structure to infinity.} since only then the number of the dualities is equal to the dimension of the manifold.
This raises a natural question: is it possible to find an alternative prescription for the cosets that reduces to the standard abelian T duality in a certain limit? In this article we will argue that while this cannot be done in the worldsheet approach, the relevant construction exists in an effective action for massless fields. In other words, we will introduce a ``modified NATD'' that can be used as a solution generating technique in the gravity approximation, just as the standard NATD and its abelian version have been used in the past. In contrast to those, the modified duality does not have a natural extension to stringy excitation. 

Another solution generating technique closely related to T duality is the so-called TsT transformation \cite{TsTdef}, which has been used to produce geometries dual to non--commutative gauge theories \cite{TsTncW1,TsTncW2} and integrable deformations of string theories on AdS$_p\times$S$^q$ \cite{TsTbetaW1,TsTbetaW2,TsTdef,BetaDefW1,BetaDefW2,BetaDefW3,BetaDefW4,BetaDefW5}. In the abelian case, when the manifold has a $U(1)\times U(1)$ symmetry, the TsT transformation consists of dualization along the first group, a coordinate transformation in the resulting solution that involves mixing the two cyclic coordinates, and dualization back \cite{TsToldW1,TsToldW2,TsTncW1,TsTncW2,TsTbetaW1,TsTbetaW2}. The key feature of this construction, which gives a family of solutions parameterized by one continuous number, is that the $U(1)\times U(1)$ symmetry is preserved at every step of the procedure. Extension of this construction to the non--abelian case runs into a challenge: NATD is known to destroy the continuous symmetries of the original background, so neither the shift nor the dualization back can be performed. One approach to addressing this challenge was recently explored in \cite{LStSt}, where a non--abelian shift was done {\it before} performing the T duality. This led to several examples of string theory solutions depending on one deformation parameter\footnote{Application of this non--abelian shift procedure to so-called bubbling geometries \cite{LLM} dual to $1/2$--BPS states in $N=4$ SYM also had an interesting interpretation in terms of the spectral flow in the dual field theory \cite{SchwSeib,SpecFlowW1,SpecFlowW2}. We refer to \cite{LStSt} for details.}, but when this parameter was switched off, the geometry described a T--dual of a group manifold, not the group itself. In this article we will address a complementary problem of constructing the TsT deformations of spaces involving products of two $n$--spheres, which recover $S^n\times S^n$ when the deformation parameter is switched off. Due to the challenge associated with destruction of symmetries by NATD, our construction will not rely on the worldsheet techniques (in contrast to the approach presented in \cite{LStSt}), but rather will be carried out in the supergravity approximation. Specifically, we will impose the symmetries expected to be preserved  by the ``non--abelian TsT deformation'' and find the most general solution of supergravity within this class. This will allow us to avoid explicit dualization and challenges associated with the second NATD. The resulting ``non--abelian TsT deformation'' turns out to be unique, and it satisfies the correct equations for the NS--NS sector of string theory in the low--energy approximation. Therefore, in contrast to the ``modified NATD'', the ``non--abelian TsT deformation'' might be extendable to the full string theory, but exploration of this possibility is beyond the scope of our article.

\bigskip
Studies of the non--abelian T--duality (NATD) have a long history \cite{NATD80w1,NATD80w2,OldNATDw1,OldNATDw2,OldNATDw3,OldNATDw4,
OldNATD1W1,OldNATD1W2,OldNATD1W3,OldNATD1W4}, but they have intensified in the last decade as this duality has been applied to produce novel backgrounds relevant for the AdS/CFT correspondence \cite{NATD101,NATDAdSCFTw1,NATDAdSCFTw2,NATDAdSCFTw3,
NATDAdSCFTw4,NATDAdSCFTw5,NATDAdSCFTw6,NATDAdSCFTw7,NATDAdSCFTw8,
NATDAdSCFTw9,NATDAdSCFTw10,NATDAdSCFTw11,NATDAdSCFTw12,ILCS12,
SfetRRw1,SfetRRw2,
NATDappW1,NATDappW2,NATDappW3,NATDappW4,NATDappW5}. There are two distinct procedures for performing NATD: one is applicable to group manifolds \cite{NTDgrp,1012}, and the other is used to dualize along cosets \cite{NTDcosetW1,NTDcosetW2,NTDcosetW3}.  Both methods have been used in the past to generate new solutions by dualizing backgrounds like $AdS_3\times S^3\times T^4$ and $AdS_5\times S^5$ along various spheres \cite{NATDAdSCFTw1,NATDAdSCFTw2,NATDAdSCFTw3,
NATDAdSCFTw4,NATDAdSCFTw5,NATDAdSCFTw6,NATDAdSCFTw7,NATDAdSCFTw8,
NATDAdSCFTw9,NATDAdSCFTw10,NATDAdSCFTw11,NATDAdSCFTw12,NATD101} and to interpret the results as interesting new solutions in M-theory \cite{TDUpliftW1,TDUpliftW2}. Since any $n$--sphere can be viewed as the $SO(n+1)/SO(n)$ manifold, most of these dualizations relied on the coset prescription, with the exception of the three--sphere, which can be alternatively viewed as the $SU(2)$ group manifold. The relationship between the coset and group dualizations of the three--sphere was explored in \cite{LStSt}, and it will be briefly reviewed in section \ref{SecSubNATDstand}. Here we just point out that in the decompactification limit, when the sphere degenerates into $\mathbb{R}^3$, it is the group prescription that reduces to the abelian T dualities along three flat directions. In section \ref{SecNATD} we will propose an analog of such group prescription for all spheres $S^n$ by studying effective gravitational actions rather than worldsheet theory. 

While a single T duality may serve as a solution generating technique by mapping two very different metrics into each other, the situation becomes even more interesting when one combines multiple dualities with coordinate transformations. Even in the abelian case, geometries with $[U(1)]^n$ isometries form representations of the $O(n,n)$ duality group \cite{OddGrpW1,OddGrpW2,OddGrpW3,OddGrpW4,OddGrpW5}. From the string theory perspective, the metrics related by $O(n,n,\mathbb{Z})$ are equivalent to each other, while the entire $O(n,n)$ with real coefficients changes the periodic identification on tori and thus produces genuinely new solutions. The solution generating techniques based on $O(n,n)$ transformations have been extensively explored in the past to produce geometries with several continuous parameters \cite{SenCvetW1,SenCvetW2,SenCvetW3,SenCvetW4,SenCvetW5,SfetRRw0}\footnote{In the heterotic case, one can also construct new geometries using a larger duality group $O(n,n+16)$ \cite{SenCvetW1,SenCvetW2,SenCvetW3,SenCvetW4,SenCvetW5}.}, and such transformations are naturally studied in the framework of the Double Field Theory \cite{OddGrpW1,OddGrpW2,OddGrpW3,OddGrpW4,OddGrpW5,DFT17w1,DFT17w2,DFT17w3}. This brings up an interesting question of extending the DFT formalism and related solution generating techniques to the case of non--abelian isometries. In the abelian case, the simplest nontrivial setting for DFT involves $O(2,2)$ which corresponds to two commuting $U(1)$ isometries. In this case, the procedure for generating solutions is known as the TsT 
transformation \cite{TsTdef}, and it amounts to dualizing along the first $U(1)$ group, performing a coordinate transformation in the resulting solution by mixing the two cyclic coordinates, and dualizing back \cite{TsToldW1,TsToldW2,TsTncW1,TsTncW2,TsTbetaW1,TsTbetaW2,TsTdef}. A non--abelian counterpart of this procedure would involve starting with a geometry with $G_1\times G_2$ symmetry, dualizing along $G_1$, mixing coordinates, and dualizing back. As explained earlier, the worldsheet version of this prescription faces a challenge associated with non--invertible nature of non--abelian T duality, so in section \ref{SecTsT} we will explore an alternative approach that works in the gravity approximation. 

\bigskip

This paper has the following organization. In section \ref{SecNATD} we propose the unique gravitational construction of geometries which have all expected symmetries of NATD applied to spheres and reduce to standard abelian duals in the decompactification limit. Specifically, in section \ref{SecSubNATDstand} we review the traditional worldsheet NATD for $S^n$, which treats this space as a coset and can be interpreted as $n(n+1)/2$ dualities when the radius of the sphere becomes large. We draw the contrast between this construction and dualization of $S^3$ as a group manifold, which reduces to three abelian dualities in the decompactification limit. In section \ref{SecSubNATDnew} we argue that a natural extension of the group procedure to other spheres 
$S^n$ in gravity involves an $n$--form field strength which takes the system outside of the NS--NS section of string theory, but which disappears in the decompactification limit. Several examples of geometries obtained by such ``gravitational NATD'' are constructed by solving equations of motion for gravity coupled to the $n$--form and a dilaton. In section \ref{SecTsT} we analyze a different but closely related problem of extending the TsT transformations to the non--abelian case. In contrast to a single NATD, no fields beyond those in the NS--NS sector of supergravity are introduced, and by solving equations of motion we propose continuous ``non--abelian TsT deformations''  for geometries containing $S^n\times S^n$ and $S^n\times T^n$ factors. Due to peculiar properties of the three--dimensional sphere, the discussion is separated into $n=3$ and $n\ne 3$ cases in sections \ref{SecSubS3S3}--\ref{SecSubS3T3} and \ref{SecSubSnTn}--\ref{SecSubSnSn}. Some technical details are presented in appendices.

\section{Gravitational version of NATD}
\label{SecNATD}

In this section we will explore a discrete transformation of gravity solutions inspired by the non--abelian T duality and apply it to geometries containing an $n$--dimensional sphere. When $n=3$, this transformation reduces to NATD along the sphere viewed as a group manifold, but in other dimensions we will get new solutions in gravity which do not have a clear interpretation on the worldsheet. Specifically, the symmetries expected for the dual uniquely fix the geometry which turns out to be supported by a non--trivial $n$--form field strength. Therefore, the ``dual solution'' cannot be embedded in the NS--NS sector of string theory unless $n=3$. In spite of this shortcoming, the result has very interesting properties as a pure gravitational solution. We will begin with reviewing the standard procedure for dualizing geometries containing an $n$--dimensional sphere and discussing shortcomings of this operation for $n\ne 3$ in section \ref{SecSubNATDstand}. Then in section \ref{SecSubNATDnew} we will propose a new procedure, demonstrate its uniqueness, and discuss the properties of the resulting gravitational solutions.

\subsection{Standard NATD for $S^n$}
\label{SecSubNATDstand}

Let us briefly review the non--abelian T duality for geometries containing the $n$--dimensional sphere and outline some shortcomings of this procedure. Specifically, we will look at the decompactification limit and demonstrate that the expected result for the duality along $T^n$ is recovered only for $n=3$. This problem will serve as a motivation for an alternative dualization procedure which will be explored in the next subsection. 

\bigskip

There are two different prescriptions for the non--abelian T duality (NATD): one works for group manifolds \cite{NTDgrp}, and the other one is applicable to cosets \cite{NTDcosetW1,NTDcosetW2,NTDcosetW3}. The latter option can be applied to any sphere $S^n$ once it is viewed as the $SO(n+1)/SO(n)$ coset. For $S^3$ one can apply both prescriptions, but the outcomes are quite different, and the relation between them was explored in \cite{LStSt}. Let us briefly review these results.

Starting with a geometry described by a pure metric with an $S^3$ factor,
\bea\label{S3start}
ds^2=ds_\perp^2+A d\Omega_3^2,
\eea
and performing the NATD along the three--sphere viewed as a group manifold, one finds the metric in the string frame as well as the Kalb--Ramond field and the dilaton
\bea\label{NatdS3}
ds_S^2&=&ds_\perp^2+\frac{1}{4}\left[\frac{dr^2}{A}+\frac{r^2 A}{A^2+r^2}d\Omega_2^2\right],\nn
B&=&\frac{1}{4}\frac{r^3}{A^2+r^2}d\Omega_2,\quad e^{2\phi}=\frac{64}{A(A^2+r^2)}\,.
\eea
In the decompactification limit, where $A$ becomes large\footnote{Specific implementations of this limit will be discussed below. Here we just write $A={\tilde A}/\eps^2$ and formally send $\eps$ to zero while keeping ${\tilde A}$ fixed.}, the transition from (\ref{S3start}) to (\ref{NatdS3}) can be summarized as
\bea\label{NatdS3grp}
ds^2=ds_\perp^2+{\tilde A} dy_a dy_a\ \rightarrow\ 
ds_S^2=ds_\perp^2+\frac{1}{4{\tilde A}}\left[{dr^2}+r^2d\Omega_2^2\right],\quad 
e^{2\phi}=\frac{64}{{\tilde A}^3},\quad B=0.
\eea
This map describes the abelian dualities along three directions spanned by $y_a$, so the geometry (\ref{NatdS3}) is the most natural counterpart of the abelian T dualities applied to $S^3$ as a three--dimensional manifold.

\begin{figure}
\begin{center}
\[
\begin{tikzpicture}[scale=5]
\node (A) at (0.575,0.575) {A};
\node (F) at (0,0) {$S^3$};
\node (D) at (1.15,0) {B};
\draw[->] (F) -- (A) node[midway,left]{\scriptsize{NATD along $SU(2)$}};
\draw[->] (A) -- (D) node[midway,right]{\scriptsize{NATD along $SO(3)/SO(2)$}};
\draw[->] (F) -- (D) node[midway,below]{\scriptsize{~~NATD along $SO(4)/SO(3)$}};
\end{tikzpicture}
\]
\end{center}
\caption{Network of non--abelian T dualities for $S^3$.}
  \label{FigS3Trngl}
\end{figure}
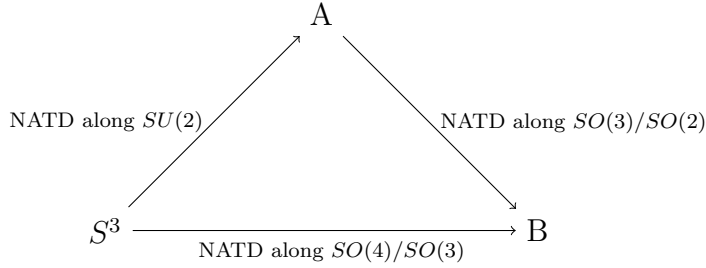

Alternatively, the NATD along the three--sphere viewed as the $SO(4)/SO(3)$ coset yields \cite{NTDcosetW1,NTDcosetW2,NTDcosetW3,LStSt}
\bea\label{NatdS3coset}
d{s}^2&=&ds_7^2+\frac{dr^2}{4A}
+\frac{Ar^2}{4(A^2+r^2)}\left\{\frac{dy_2^2}{y_1^2}+\bigg[dy_1\frac{(A^2+r^2)}{Ar^2}+dy_2\frac{r^3+(A^2+r^2)y_2}{Ar^2y_1}\bigg]^2\right\}\,,\nn
e^{2\phi}&=&\frac{64}{(Ary_1)^2},\quad B=0\,.
\eea
As demonstrated in \cite{LStSt}, geometry (\ref{NatdS3coset}) can be obtained from (\ref{NatdS3})  by an additional dualization along $S^2$ viewed as the $SO(3)/SO(2)$ coset (see figure \ref{FigS3Trngl}). Taking decompactification limits, one can argue that the abelian version of the transition from (\ref{S3start}) to (\ref{NatdS3coset}) describes dualities along six directions. Furthermore, while the duality (\ref{NatdS3coset}) preserves the chirality of worldsheet fermions, the prescription (\ref{NatdS3grp}) changes it, therefore the second option maps IIB string theory into type IIA and vice versa. 

In contrast to the three--dimensional case, to perform duality along other spheres $S^n$, one must view them as  the $SO(n+1)/SO(n)$ manifolds, and the coset prescription effectively dualizes along $SO(n+1)$. Specifically, one can start with a sigma model that has a metric and a vanishing Kalb--Ramond field,
\bea\label{SnStart}
ds^2=ds_\perp^2+A d\Omega_n^2=ds_\perp^2+A (V^T dg^T dg V),\quad B=0,
\eea
where $g\in SO(n+1)$ and $V$ is a constant vector in $(n+1)$ dimensions, which can be chosen to have components 
$V_k=\delta^1_k$. The warp factor $A$ may depend on the transverse coordinates, and some Ramond--Ramond fluxes may be present. Dualization of this geometry along $g$ directions leads to a sigma model containing a metric and a $B$ field given by \cite{LStSt}
\bea\label{CosetTwist3}
S_{\text{dual}}=\int d^2\sigma (\mathcal{Y}^{-1})_{ab}\d_+u^a \d_-u^b\,.
\eea
Here
\bea\label{CosetTwist4}
\mathcal{Y}^{ab}=A(V^T t^at^bV)+{f_{ab}}^c u^c\,,
\eea
$t^a$ are generators of $SO(n+1)$ with structure constants ${f_{ab}}^c$, and $u^a$ are the $n$ dual coordinates corresponding to generators of $SO(n+1)/SO(n)$. Note that the geometry (\ref{CosetTwist3}) has no isometries. 

\begin{figure} 
\ \vskip -0.4cm
\[
\begin{tikzpicture}[scale=5]
\node (A) at (0.575,0.575) {A};
\node (F) at (0,0) {$S^n$};
\node (D) at (1.15,0) {B};
\draw[->] (F) -- (A) node[midway,left]{\scriptsize{}};
\draw[->] (A) -- (D) node[midway,right]{\scriptsize{}};
\draw[->] (F) -- (D) node[midway,below]{\scriptsize{~~NATD along $SO(n+1)/SO(n)$}};
\node (X) at (2.075,0.575) {A};
\node (Y) at (1.5,0) {$T^n$};
\node (Z) at (2.65,0) {B};
\node (H) at (2.6,0.47) {\scriptsize{NATD along}};
\node (H) at (2.6,0.38) {\scriptsize{$SO(n)/SO(n-1)$}};
\draw[->] (Y) -- (X) node[midway,left]{\scriptsize{T-dual\ \ }};
\draw[->] (X) -- (Z) node[midway,right]{};
\draw[->] (Y) -- (Z) node[midway,below]{\scriptsize{\phantom{AAA}}};
\node (W) at (0.575,-0.2) {(a)};
\node (W) at (2.075,-0.2) {(b)};
\end{tikzpicture}
\]
\caption{Network of non--abelian T dualities for $S^n$ and $T^n$. The goal of section \ref{SecSubNATDnew} is to construct the geometry corresponding to point A in figure (a). Figure (b) illustrates the dualities involved in the decompactification limit of this solution.}
  \label{Fig2Trngl}
\end{figure}
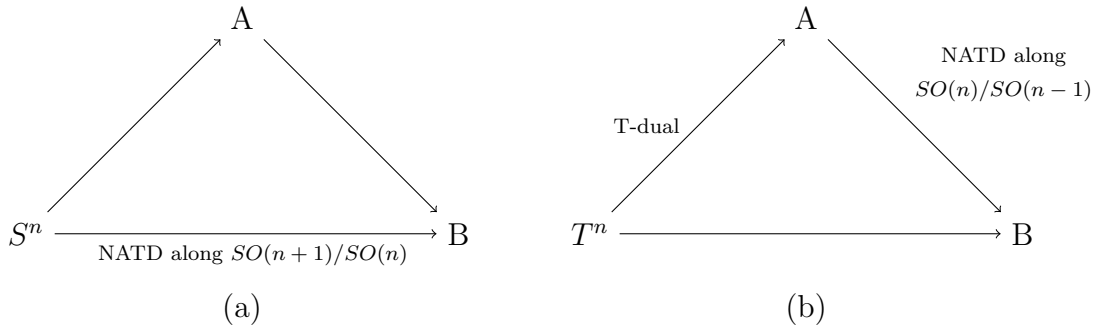

The procedure outlined above implies that the worldsheet techniques effectively dualize along $\frac{n(n+1)}{2}$ directions \cite{LStSt}. In particular, regardless of the dimension of the sphere, the coset prescription always maps IIB string theory into itself, the fact that becomes very important when one explores backgrounds with Ramond--Ramond fields \cite{SfetRRw1,SfetRRw2,NATDappW1,NATDappW2,NATDappW3,NATDappW4,NATDappW5,LStSt}. This raises a natural question about a counterpart of the solution (\ref{NatdS3}) for spheres $S^n$ with $n\ne 3$, i.e. about the point A in the Figure \ref{Fig2Trngl}(a). Unfortunately, it is not clear how to find the relevant solutions by performing algebraic procedures with sigma model, so in the next subsection we will pursue an alternative path by imposing an ansatz inspired by (\ref{NatdS3}) and solving the relevant equations of motion. 

\subsection{New procedure in gravity}
\label{SecSubNATDnew}

In the previous subsection we discussed the standard non--abelian dualization of the metric (\ref{SnStart}) and observed that this procedure corresponds to performing $\frac{n(n+1)}{2}$ dualities. In the case of $S^3$, such dualization can be performed in two steps, first going from (\ref{S3start}) to (\ref{NatdS3}) by performing three dualities, and then transforming (\ref{NatdS3}) into (\ref{NatdS3coset}) by performing three more, so one may hope to have a counterpart of the geometry (\ref{NatdS3}) as the point A in the Figure \ref{Fig2Trngl}(a). If such a point exists, it is expected to have an $SO(n)$ isometry, so the geometry supported by a dilaton and a flux must have the form\footnote{In this subsection metrics are always written in the Einstein frame.} 
\bea\label{NatdSn}
ds^2&=&e^\psi\left[ds_\perp^2+\frac{1}{A}\left(f_1 dr^2+f_2 r^2 d\Omega_{n-1}^2\right)\right]\nn
e^{2\phi}&=&e^{-\frac{d-2}{2}\psi},\quad C^{(n-1)}=\frac{\sqrt{\beta} r^n f_4}{A^{\frac{n+1}{2}}}  d\Omega_{n-1};
\quad e^\psi=A^p f_3,\quad p=\frac{2n}{d-2}
\eea
Here $(f_1,f_2,f_3,f_4)$ are functions of $r$ and $A$. Taking the decompactification limit similar to (\ref{NatdS3grp}) and setting all functions $f_k$ to one, we arrive at the geometry
\bea\label{NatdSnTor}
ds^2&=&e^{\psi}\left[ds_\perp^2+\frac{1}{\bar A}\left[dr^2+r^2 d\Omega_{n-1}^2\right]\right],\quad
e^{2\phi}=\frac{1}{{\bar A}^n},\quad C^{(n-1)}=0\,,
\eea
which is a T dual of a pure metric along ${\mathbb{R}}^n$ spanned by $S^{n-1}$ and the $r$ coordinate. This geometry corresponds to the point $A$ in figure \ref{Fig2Trngl}(b). Therefore, it is reasonable to interpret (\ref{NatdSn}) as the result of applying $n$ dualities to (\ref{SnStart}). Additional dualization of (\ref{NatdSn}) along $S^{n-1}\sim SO(n)/SO(n-1)$ reproduces the correct number of dualities between points $S^n$ and $B$ of the figure \ref{Fig2Trngl}(a):
\bea
n+\frac{n(n-1)}{2}=\frac{n(n+1)}{2}.
\eea
Note that unless $n=3$ or $C^{(n-1)}=0$, the geometry (\ref{NatdSn}) cannot be interpreted as a set of the 
NS--NS fields, so it cannot be obtained from (\ref{SnStart}) by manipulating the sigma model. Therefore, in this subsection we will focus on obtaining the solution (\ref{NatdSn}) from a supergravity-type action containing only the metric, the dilaton, and the flux.

Motivated by the NS--NS sector of string theory, we consider the action
\bea\label{Action}
S=\int d^d x \sqrt{-g}\left[R-\frac{\alpha}{2}(\nabla\psi)^2-\frac{1}{2n!}e^{\nu\psi}
F_{\alpha_1\dots \alpha_{n}}F^{\alpha_1\dots \alpha_{n}}\right]
\eea
and impose the ansatz (\ref{NatdSn}). In the decompactification limit (\ref{NatdSnTor}), the flux disappears, so (\ref{Action}) can be interpreted as the action for the metric and the dilaton in the Einstein frame. In this limit, (\ref{NatdSnTor}) is obtained by performing a T--duality of (\ref{SnStart}) along the torus $T^n$. To simplify subsequent equations, it is convenient not to set $\psi$ to be equal to the dilaton, but rather to impose the simplest relation between the metrics in the string and Einstein frames:
\bea
ds_E^2=e^{\psi}ds_S^2.
\eea
This leads to the expressions for the dilaton and for the parameter $\alpha$ in (\ref{Action}):
\bea
e^{2\Phi}=\exp\left[-\frac{d-2}{2}\psi\right],\quad \alpha=\frac{d-2}{2}\,.
\eea
If we also set $\nu=n-1$, then equations of motion for the action (\ref{Action}) become
\bea\label{EomNATD}\label{EomSn}
&&R_{\mu\nu}-\frac{1}{2}g_{\mu\nu}R=\frac{e^{(n-1)\psi}}{2(n-1)!}
\left[F_{\mu\alpha_1\dots \alpha_{n-1}}{F_\nu}^{\alpha_1\dots \alpha_{n-1}}-\frac{1}{2n}g_{\mu\nu}F_{\alpha_1\dots \alpha_{n}}{F}^{\alpha_1\dots \alpha_{n}}\right]
\nn
&&\phantom{R_{\mu\nu}-\frac{1}{2}g_{\mu\nu}R=}+
\frac{d-2}{4}\left[\d_\mu\psi\d_\nu\psi-\frac{1}{2}g_{\mu\nu} (\d\psi)^2\right]\,,\\
&&\nabla_\mu(e^{(n-1)\psi}F^{\mu\alpha_1\dots \alpha_{n-1}})=0,\quad
\nabla^2 \psi=\frac{n-1}{d-2}\frac{e^{(n-1)\psi}}{n!}F_{\alpha_1\dots \alpha_{n}}F^{\alpha_1\dots \alpha_{n}}\,.\nonumber
\eea
To find solutions of these equations beyond the limit (\ref{NatdSnTor}), we focus on a specific example of the initial metric (\ref{SnStart}):
\bea\label{SnSeps}
ds^2=ds_{1,n-1}^2+du^2+\left(\frac{1}{\eps}+u\right)^2 d\Omega_n^2.
\eea
This is flat space, and the decompactification limit is obtained by sending parameter $\eps$ to zero, while keeping $u$ fixed. Using (\ref{NatdS3}) as an inspiration, we look for the ``dual'' solution 
in the form (\ref{NatdSn}) assuming that $(f_1,f_2,f_3,f_4)$ are functions of only one variable $x$:
\bea\label{SnAfact}
x=\frac{\eps^2 r^2}{A^2},\quad A=(1+\eps u)^2\,.
\eea
General analysis of equations of motion (\ref{EomNATD}) for the 
ansatz  (\ref{NatdSn}), (\ref{SnAfact}) is presented in the Appendix \ref{AppEinstNATD}, and here we just summarize the main features of the solutions.
\begin{enumerate}[(a)]
\item In the decompactification limit, the standard T duality gives
\bea
f_1=f_2=f_3=1,
\eea
and function $f_4$ remains undetermined since flux disappears in this limit. Performing perturbative expansions in $\eps$, one finds the series
\bea\label{PertEll}
f_\ell=1+\sum a^{(\ell)}_k x^k,\quad \ell=1,2,3;\quad  f_4=\sqrt{\beta}+\sum a^{(4)}_k x^k,\quad 
\eea
where the numerical coefficients $(a^{(1)}_k,a^{(2)}_k,a^{(3)}_k,a^{(4)}_k)$ depend only on one continuous parameter $\beta$ and the number of dimensions $n$. The first few terms in the expansions are given by (\ref{PertArbN}). Although the series may have a finite radius of convergence, uniqueness of the solution persists to all values of $x$ up to possible singularities of the metric (\ref{NatdSn}). 
\item Numerical integration of the system (\ref{EomSn}) with boundary conditions (\ref{PertArbN}) indicates that the metric (\ref{NatdSn}) encounters a curvature singularity at a finite value of $x$ unless parameter $\beta$ takes a specific value $\beta={\hat \beta}_n$ given in Table \ref{Table1}. For this $\beta$, functions $f_\ell$ are defined for all positive values of $x$. Examples of functions $f_\ell$ for various values of $\beta$ and $n$ are presented below.

\item To simplify numerical integration, it is convenient to lower the order of equations. Observing that the ansatz (\ref{NatdSn}) has a residual symmetry under rescaling of function $f_3$ with fixed ``string metric'', we conclude that the order of the system can be lowered by defining a new function $f_3$ via (\ref{f3g3App}):
\bea\label{f3g3}
f_3=\exp\left[\int \frac{g_3 dx}{d-2}\right]\,.
\eea
The rescaling symmetry ensures that all integrals disappear from the Einstein's equations. Below we will present some numerical results for the profiles of functions $(f_1,f_2,f_4,g_3)$. 

\item For $\beta={\hat\beta}_n$, numerical integration yields regular functions $f_\ell$ for all values of $x$, and  the leading contributions at large $x$ for $n>3$ are\footnote{Some subleading corrections are given by (\ref{InfExpandApp}).}
\bea\label{funcAsymp}
f_1=\frac{c_1}{\sqrt{y}},\quad f_2=\frac{c_2}{\sqrt{y}},\quad f_4=c_4 y^{-\frac{n+1}{4}},\quad
f_3=\exp\left[\frac{n}{d-2}\ln y\right]
\eea
Here
\bea\label{funcAsympC}
c_1=\frac{c_2 \left[9n-2+c_4^2c_2^{1-n} (n-1)\right]}{4 (n-2)}\,,
\eea
and the values of $(c_2,c_4)$ are determined by numerical integration. Note that this asymptotic behavior is not encountered for $n=2,3$. In the former case, ${\hat\beta}_3=2$ leads to the exact solution
\bea\label{n3Exact}
n=3:\quad f_1=1,\quad f_2=f_3=\frac{1}{1+4x},\quad f_4=\frac{2}{1+4x}\,
\eea
corresponding to the non--abelian dual (\ref{NatdS3}) of flat space along $S^3$ viewed as a group.
In the $n=2$ case, the asymptotic behavior of functions $f_\ell$ for ${\hat\beta}_2=5.11$ is rather complicated, and we will not discuss it further.

The leading contribution to the geometry (\ref{NatdSn}) from the asymptotic expansions (\ref{funcAsymp}) is
\bea\label{S4asympInf}
ds^2&=&ds_\perp^2+\frac{1}{\eps r}\left[c_1 dr^2+r^2 c_2 d\Omega_{n-1}^2\right],\nn
e^{2\phi}&=&\frac{1}{A^n}f_3,\quad 
C^{(n-1)}=\frac{r^n}{A^{p}} f_4 d\Omega_{n-1}\rightarrow 
c_4 (r/\eps)^{\frac{n+1}{2}}d\Omega_{n-1}
\eea
\end{enumerate}
\begin{table}
\begin{center}
\begin{tabular}{|c|c|c|c|c|c|c|c|c|c|c|c|c|}
\hline
$n$&2&3&4&5&6&7&8&9&10&11&12\\
\hline
${\hat\beta}_n$&5.11&2&1.25&1&0.83&0.71&0.52&0.5&0.47&0.45&0.36\\
\hline
\end{tabular}
\end{center}
\caption{The values of $\beta$ leading to regular geometries for the duals of spheres $S^n$.}
\label{Table1}
\end{table}
Since qualitative behavior of functions $f_\ell$ is the same for all $n>3$, let us present some details for $n=2,3,4$. 
\begin{itemize}
\item {${\bf n=3}$}

Expanding solutions of the system (\ref{EomSn}) in powers of $x$, one finds 
\bea\label{f1234ExpS3main}
f_1&=&1-\frac{9{\bar\beta}x}{35}+\frac{3{\bar\beta}(3920+3307{\bar\beta})x^2}{53900}+\mathcal{O}(x^3)\nn
\frac{1}{f_2}&=&1+4x+\frac{181{\bar\beta}x}{70}-\frac{{\bar\beta}(67270+32313{\bar\beta})x^2}{13475}+\mathcal{O}(x^3),
\eea
as well as similar expansions for $(f_3,f_4)$ given by (\ref{f1234ExpS3}). Here we defined 
${\bar\beta}=\beta^2-2$. The series for  $(f_1,f_2^{-1},f_3^{-1},f_4^{-1})$ truncate for 
$\hat\beta_3=\sqrt{2}$ leading to the exact solution (\ref{n3Exact}), which describes NATD along the three--dimensional sphere viewed as a  group manifold (\ref{NatdS3}). Profiles of various functions for this case are shown in Figure \ref{FigN3}(a). \\
For $\beta<\hat\beta_3$, series (\ref{f1234ExpS3main}), (\ref{f1234ExpS3}) can be used as a starting point for numerical integration, which shows that the geometry (\ref{NatdSn}) encounters a curvature singularity at some finite value of $x$, where functions $(f_1,f_2)$ go to zero. Typical profiles of various functions in this phase are presented in figure \ref{FigN3}(b).\\ 
For $\beta>\hat\beta_3$, series numerical integration of the system again leads to a curvature for the geometry (\ref{NatdSn}) where functions $(\frac{1}{f_1},f_2)$ go to zero. Typical profiles of various functions in this phase are presented in figure \ref{FigN3}(c).

\begin{figure}
\begin{tabular}{llcl}
\begin{tabular}{c} 
(a)\\ \ \\ \\ \ \\ \\ \ \\ \ \\
\end{tabular}
&\hskip -0.6cm \ 
  \includegraphics[width=0.42 \textwidth]{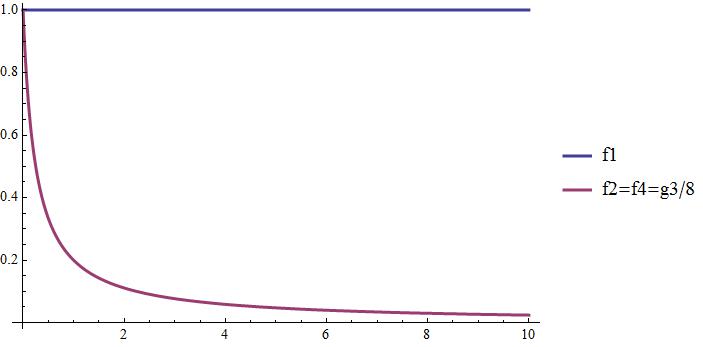}&\ &
 \\
\begin{tabular}{c} 
(b)\\ \ \\ \\ \ \\ \\ \ \\ \ \\
\end{tabular}
&\hskip -0.6cm \ 
  \includegraphics[width=0.42 \textwidth]{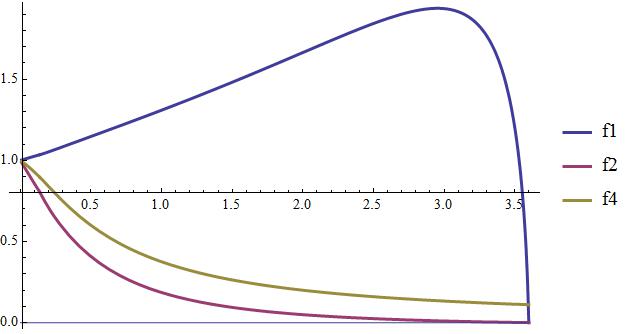}&\ &
  \hskip -0.8cm \ 
   \includegraphics[width=0.47 \textwidth]{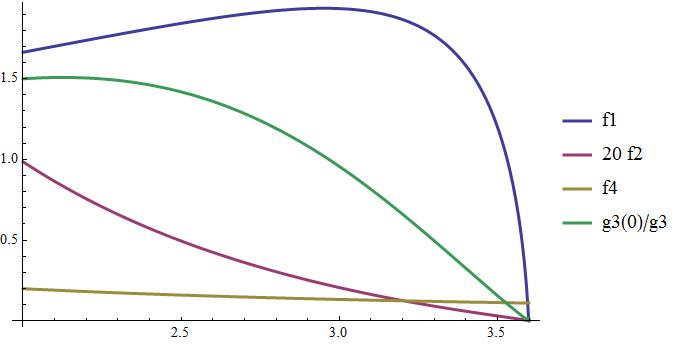}\\
\begin{tabular}{c} 
(c)\\ \ \\ \\ \ \\ \\ \ \\ \ \\
\end{tabular}
&\hskip -0.6cm \ 
  \includegraphics[width=0.42 \textwidth]{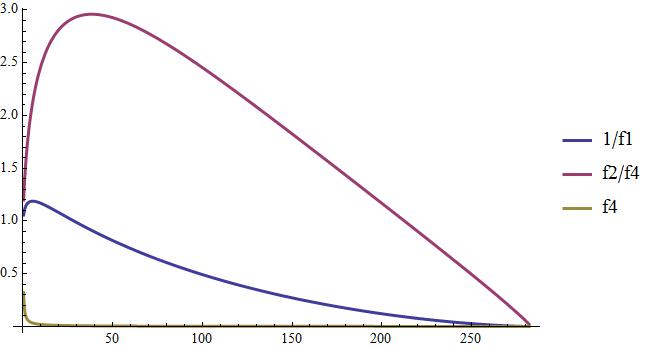}&\ \ &
  \hskip -0.8cm \ 
   \includegraphics[width=0.47 \textwidth]{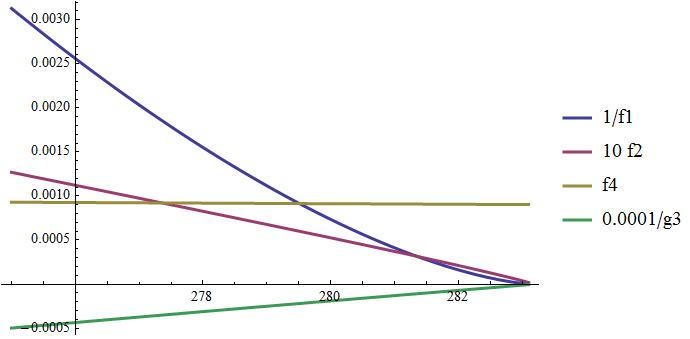}
\end{tabular}
\caption{Profiles of various functions appearing in the ansatz (\ref{NatdSn}) for the dual along $S^3$: (a) the standard NA T--dual with $\beta={\hat\beta}_3={2}$; (b) solution with $\beta=1<{\hat\beta}_3$; (c) solution with $\beta=3>{\hat\beta}_3$.
}
\label{FigN3}
\end{figure}

\begin{figure}
\begin{tabular}{llcl}
\begin{tabular}{c} 
(a)\\ \ \\ \\ \ \\ \\ \ \\ \ \\
\end{tabular}
&\hskip -0.6cm \ 
  \includegraphics[width=0.42 \textwidth]{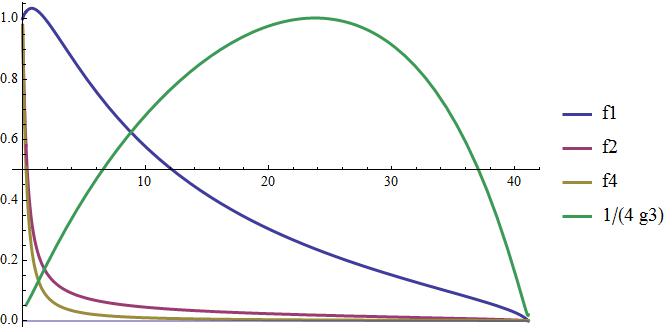}&\ &
  \hskip -0.8cm \ 
   \includegraphics[width=0.47 \textwidth]{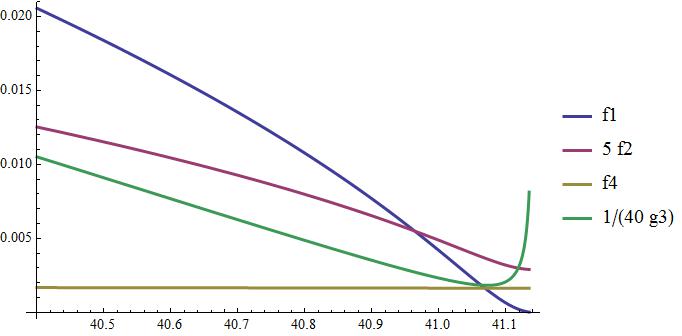}\\
  \begin{tabular}{c} 
(b)\\ \ \\ \\ \ \\ \\ \ \\ \ \\
\end{tabular}
&\hskip -0.6cm \ 
  \includegraphics[width=0.42 \textwidth]{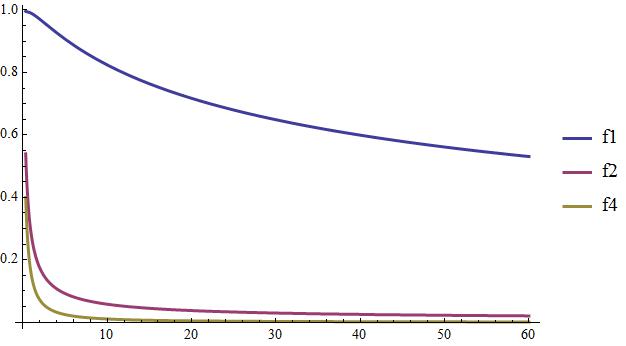}&\ &
  \hskip -0.8cm \ 
   \includegraphics[width=0.47 \textwidth]{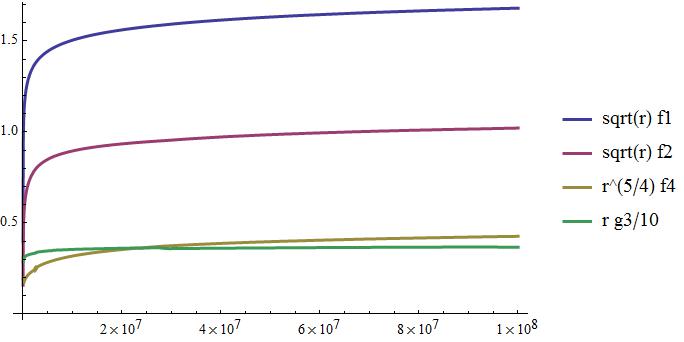}\\ 
\begin{tabular}{c} 
(c)\\ \ \\ \\ \ \\ \\ \ \\ \ \\
\end{tabular}
&\hskip -0.6cm \ 
  \includegraphics[width=0.42 \textwidth]{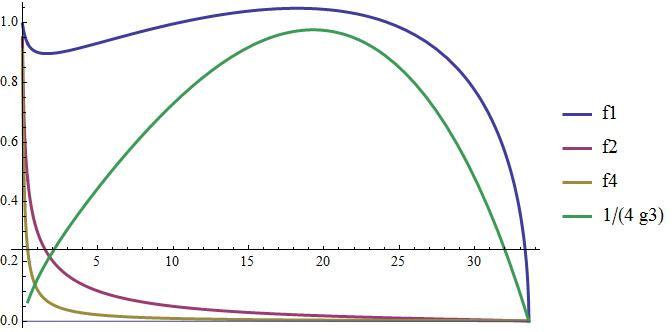}&\ \ &
  \hskip -0.8cm \ 
   \includegraphics[width=0.47 \textwidth]{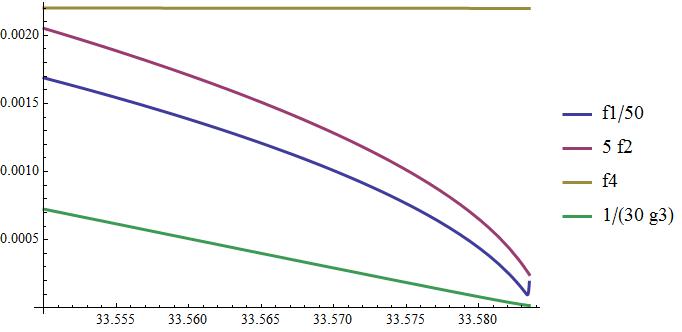}
\end{tabular}
\caption{Profiles of various functions appearing in the ansatz (\ref{NatdSn}) for the dual along $S^4$: (a) $\beta=1<{\hat\beta}_4$; (b) $\beta={\hat\beta}_4=5/4$; (c) 
$\beta=2>{\hat\beta}_4$.
}
\label{FigN4}
\end{figure}

\item {${\bf n\ge 4}$}

Expanding solutions of the system (\ref{EomSn}) in powers of $x$, one finds the series (\ref{PertArbN}), the higher dimensional counterparts of (\ref{f1234ExpS3main}). In contrast to 
expansions (\ref{f1234ExpS3main}) which truncate for $\beta=2$, there seem to be no truncations or recognizable patters in higher dimensions regardless of the value of $\beta$. Therefore, one has to resort to numerical integration using series (\ref{PertArbN}) as boundary conditions. This leads to unique solutions for any pair of values $(n,\beta)$, but as in the $n=3$ case,  the geometries develop naked singularities unless $\beta$ is tuned to a specific value 
${\hat\beta}_n$ given in Table \ref{Table1}. Functions $(f_1,f_2,g_3,f_4)$ have the same qualitative features for all $n\ge 4$, so in figure \ref{FigN4} we present the profiles for $n=4$. \\
For $\beta<\hat\beta_4$, the geometry (\ref{NatdSn}) encounters a curvature singularity at some finite value of $x$, where functions $(f_1,f_2)$ go to zero. Typical profiles of various functions in this phase are presented in figure \ref{FigN4}(a).\\ 
For $\beta=\hat\beta_4$, functions $(f_1,f_2,g_3,f_4)$ can be integrated to infinity, where their asymptotic behavior is given by (\ref{funcAsymp}). The resulting geometry gives the gravitational counterpart of the non--abelian T duality which corresponds to point A in figure \ref{Fig2Trngl}(a). Typical profiles of various functions for this solution are presented in figure \ref{FigN4}(b).\\ 
For $\beta>\hat\beta_4$, numerical integration of the system again leads to a curvature singularity for the geometry (\ref{NatdSn}) where functions $(\frac{1}{f_1},f_2)$ go to zero. Typical profiles of various functions in this phase are presented in figure \ref{FigN4}(c).

\item {${\bf n=2}$}

The phases for $n=2$ are similar to those for $n\ge 4$ with two qualitative differences: 
function $f_2$ typically goes through zero at some $x=x_0$ without generating a singularity, and asymptotic behavior at large $x$ for 
$\beta={\hat\beta}_2$ differs from (\ref{funcAsymp})-- (\ref{funcAsympC}). \\
\ \\
Specifically, for a large range of $\beta$, which contains the critical value 
$\beta={\hat\beta}_2$, function $f_2$ goes through zero at some $x=x_0$, while $(f_1,g_3,f_4)$ remain finite. Near this point one finds
\bea
f_2\simeq a(x-x_0)^2=\frac{a\eps^4}{A^4}(r^2-r_0^2)^2,\quad r_0=\frac{A\sqrt{x_0}}{\eps},
\eea 
so the geometry remains regular since one can introduce a new coordinate ${\bar r}$:
\bea
f_2 dr^2\simeq \frac{a\eps^4}{4A^4}(r+r_0)^2 d{\bar r}^2
\simeq b d{\bar r}^2,\quad {\bar r}=(r-r_0)^2.
\eea
Typical profiles of various functions illustrating this feature are presented in figure \ref{FigN2}. Once the solution is continued beyond $x=x_0$, it either extends to infinity or ends in a singularity in a manner similar to the $n\ge 4$ with minor quantitative differences in the asymptotic expansions analogous to (\ref{funcAsymp})-- (\ref{funcAsympC}):\\
For $\beta\ne{\hat\beta}_n$ numerical integration of the system leads to a curvature singularities for the geometry (\ref{NatdSn}) where either $({f_1},f_2)$ or $(\frac{1}{f_1},f_2)$ go to zero. Typical profiles of various functions in this phase near the singularity are similar to those presented in figure \ref{FigN4}(a,c).\\
For $\beta={\hat\beta}_2$, functions $(f_1,f_2,g_3,f_4)$ remain finite for all $x>x_0$ and solution extends to infinity in a manner similar to the one found in the $n\ge 4$ case.

\begin{figure}
\begin{center}
  \includegraphics[width=0.7 \textwidth]{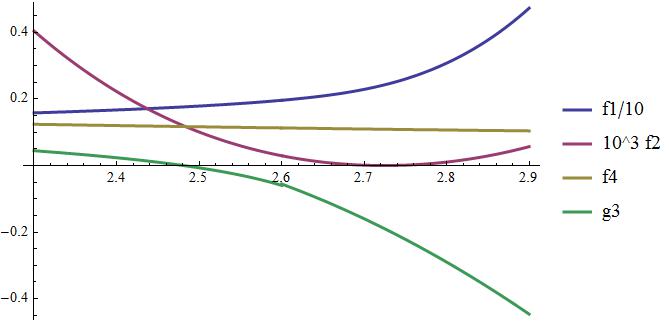}
\end{center}
\caption{A peculiar feature of the solution (\ref{NatdSn}) for $n=2$: for a wide range of $\beta$ around ${\hat\beta}_2=5.11$, function $f_2$ bounces off zero before the solution terminates as in figures \ref{FigN4}(a,c) for $\beta\ne {\hat\beta}_2$ or continues to infinity as in figure \ref{FigN4}(b) for $\beta= {\hat\beta}_2$. The geometry remains regular at the location of the bounce.
}
\label{FigN2}
\end{figure}

\end{itemize}

\noindent
To summarize, in this section we proposed a gravity counterpart of NATD that extends the well--known result (\ref{NatdS3}) from $S^3$ to spheres with other dimension. In contrast to the coset prescription that effectively leads to $\frac{n(n+1)}{2}$ dualities \cite{LStSt}, our construction reduces to $n$ abelian T dualities in the decompactification limit, where $S^n$ is replaced by a torus. This feature comes at the expense of introduction of a gauge potential $C^{(n-1)}$, which is not a part of the NS--NS sigma model unless $n=3$, thus ruling out interpretation of our solution as a string theory duality. Nevertheless, viewed from the perspective of a gravity action (\ref{Action}), our {\it unique} solutions appear to be on the same footing for all values of $n$, and it would be interesting to find a physical reason underlying such uniformity. Our main result is summarized by (\ref{NatdSn}), (\ref{PertEll}) with values of $\beta$ given in Table \ref{Table1} and explicit series given by (\ref{PertArbN}). Unfortunately, for finite values of $r$ we had to rely in numerical integration to obtain the profiles presented in figures \ref{FigN4}(b), \ref{FigN2}(b), but asymptotic geometry (\ref{S4asympInf}) at large $r$ becomes very simple.

\section{Non--abelian TsT transformations}
\label{SecTsT}

In the previous section we constructed the gravity solutions which extended the NATD of $S^3$ as a group manifold to other spheres. T--duality transformations, either abelian of non--abelian, map two geometries into each other, but to introduce families  of solutions specified by continuous parameters one needs to combine dualities with coordinate transformations. In the abelian case, such combinations give rise to U--duality groups \cite{SenCvetW1,SenCvetW2,SenCvetW3,SenCvetW4,SenCvetW5,OddGrpW1,OddGrpW2,OddGrpW3,
OddGrpW4,OddGrpW5}\footnote{These groups often contain S--duality as well, but we will focus on the transformations involving only T--dualities and shifts.}, which have been used to generate charged black holes \cite{SenCvetW1,SenCvetW2,SenCvetW3,SenCvetW4,SenCvetW5}, geometries dual to non--commutative gauge theories \cite{TsTncW1,TsTncW2}, and integrable deformations of string theories on AdS$_p\times$S$^q$ \cite{TsTbetaW1,TsTbetaW2}. The simplest U--duality group is encountered for geometries with $U(1)\times U(1)$ isometries, and in this case the solution generating technique is called the TsT transformation \cite{TsTdef}, and it consists of a T duality, a coordinate transformation, and a T duality back to the original frame. The goal of this section is to generalize the TsT transformations from spaces with $S^1\times S^1$ factors to geometries containing the product $S^n\times S^n$. 

We will begin with reviewing the standard TsT deformations, challenges associated with extending them to the non--abelian case, and previous progress in this direction. In particular, we will see that the most direct non--abelian counterpart of the TsT transformation is unlikely to be constructible by algebraic manipulations using worldsheet techniques. Then in sections \ref{SecSubS3S3}--\ref{SecSubSnSn} we will propose the ``gravitational TsT'', the unique families of solutions which have the properties expected from extending TsT deformations of metrics with 
$U(1)\times U(1)$ isometries to the non--abelian case. Since the $S^3$ turns out to be very different from the other spheres, we will separate the discussion into the $S^3\times S^3$ and $S^n\times S^n$ cases (sections \ref{SecSubS3S3} and \ref{SecSubSnSn}). Closely related deformations of $S^3\times T^3$ and $S^n\times T^n$ are discussed in sections \ref{SecSubS3T3} and \ref{SecSubSnTn}.

\subsection{Review of known results}
\label{SecTsTknown}
Let us begin with reviewing the abelian version of the TsT transformation. In the simplest setting, ones starts with 
the solution of string theory described by the metric 
\bea\label{StartTsTab1}
ds^2=F dx^2+G dy^2+ds_\perp^2
\eea
and vanishing Kalb--Ramond field, performs a T duality along $x$ direction, a shift 
$y\rightarrow y+\gamma x$ and another T duality. This leads to another exact solution of string theory
\bea\label{TsTabel1}
ds^2=\frac{F dx^2+G dy^2}{1+\gamma^2 FG}+ds_\perp^2,\quad 
B=\frac{\gamma FG}{1+\gamma^2 FG}dx\wedge dy,\quad e^{2\Phi}=\frac{1}{1+\gamma^2 FG}
\eea
Here $ds_\perp^2$ is the part of the metric that does not contain coordinates $(x,y)$. If the starting point (\ref{StartTsTab1}) solved equations of motion only in the supergravity approximation, than so does the deformation (\ref{TsTabel1}). In this article we will work in such supergravity approximation.

For the subsequent discussion, it is useful to generalize the deformation (\ref{TsTabel1}) to the starting geometry with several pairs of $x$ and $y$ coordinates:
\bea\label{StartTsTab}
ds^2=F dx_a dx_a+G dy_b dy_b+ds_\perp^2,
\eea
where  indices $a$ and $b$ go from one to $n$.
Then the counterpart of the geometry (\ref{TsTabel1}) with equal deformation parameters in all $(x_k,y_k)$ planes reads
\bea\label{TsTabel}
ds^2=\frac{1}{1+\gamma^2 FG}[F dx_a dx_a+G dy_b dy_b]+ds_\perp^2,\quad 
B=\frac{\gamma FG}{1+\gamma^2 FG}dx_a\wedge  dy_a
\eea
Note that while the original geometry (\ref{StartTsTab}) has the $SO(n)\times SO(n)$ symmetry under independent rotations of $x$ and $y$ coordinates, the deformation (\ref{TsTabel}) breaks this symmetry to a diagonal $SO(n)$, as can be seen from the expression for the $B$ field. This breaking will play an important role in our discussion below.

To introduce a non--abelian counterpart of the above construction, we replace the starting point (\ref{StartTsTab}) by a geometry that contains a product $S^3\times S^3$:
\bea\label{StartTsTabSS}
ds^2=4F d\Omega_3+4G d{\tilde\Omega}_3+ds_\perp^2\,.
\eea
Since both spheres are group manifolds of $SU(2)$, they can be parameterized by
\bea
g\in SU(2),\quad {\tilde g}\in {\widetilde{SU(2)}}.
\eea
Two proceed, we define the one--forms $(\sigma_a,{\tilde\sigma}_b)$ invariant under the left actions of $SU(2)$ and $\widetilde{SU(2)}$:
\bea\label{SignaLdef}
\sigma_a=-i\,\mbox{tr}(g^{-1}dg\,t_a),\quad {\tilde\sigma}_b=-i\,\mbox{tr}({\tilde g}^{-1}d{\tilde g}\,{\tilde t}_b).
\eea
Here $t_a$ and ${\tilde t}_b$ are hermitean generators of the groups. The metric (\ref{StartTsTabSS}) can be rewritten as
\bea\label{StartTsT}
ds^2=F \sigma_a\sigma_a+G {\tilde\sigma}_b{\tilde\sigma}_b+ds_\perp^2\,,
\eea
and it is invariant under the $[SU(2)]^4$ transformations:
\bea\label{SymmSU24}
SU(2)_L\times SU(2)_R\times {\widetilde{SU(2)}}_L\times {\widetilde{SU(2)}}_R:\quad
g\rightarrow h_L g h_R,\quad {\tilde g}\rightarrow {\tilde h}_L {\tilde g} {\tilde h}_R,
\eea
While coordinate--independent matrices $(h_L,{\tilde h}_L)$ cancel in (\ref{SignaLdef}) and (\ref{SymmSU24}), matrix $h_R$ rotates index $a$ in (\ref{StartTsT}), and ${\tilde h}_R$ rotates index $b$. In the decompactification limit of the spheres,
\bea
\sigma_a\rightarrow dx_a,\quad {\tilde\sigma}_b\rightarrow dy_b,
\eea
the geometry (\ref{StartTsT}) reduces to (\ref{StartTsTab}) with $n=3$, and the symmetry (\ref{SymmSU24}) becomes 
\bea
[U(1)]^3\times SO(3)\times [{\widetilde{U(1)}}]^3\times {\widetilde{SO(3)}}\,,
\eea
in agreement with our discussion above. Then the symmetries of the geometry (\ref{TsTabel}) suggest that the deformed version of (\ref{StartTsT}) would have a reduced symmetry
\bea
SU(2)_L\times {\widetilde{SU(2)}}_L\times SU(2)_{diag}:\quad
g\rightarrow h_L g h,\quad {\tilde g}\rightarrow {\tilde h}_L {\tilde g} {h}.
\eea
The most general metric and B field with these symmetries have the form
\bea\label{SignaGeneral}
ds^2=f_1 \sigma_a\sigma_a+f_2 {\tilde\sigma}_b{\tilde\sigma}_b+
f_3\sigma_a{\tilde\sigma}_a+ds_\perp^2,
\quad B=f_4 \sigma_a\wedge {\tilde\sigma}_a\,.
\eea
There are several interesting cases of such geometries:
\begin{figure}
\[
\begin{tikzpicture}[scale=7]
\node (A) at (0,0) {$[S^3\times S^3]_n$};
\node (F) at (0.35,0.35) {$A_n$};
\node (C) at (0.35,-0.35) {$B_n$};
\node (D) at (0.7,0) {$C_n$};
\node (E) at (1.35,0) {$D_n$};
\draw[->] (A) -- (F) node[midway,left]{\scriptsize{Dual along $g$}};
\draw[->] (A) -- (C) node[midway,left]{\scriptsize{Dual along $h$}};
\draw[->] (F) -- (D) node[midway,right]{\scriptsize{Dual along $h$}};
\draw[->] (C) -- (D) node[midway,right]{\scriptsize{Dual along $g$}};
\draw[->] (D) -- (E) node[midway,below]{\scriptsize{Dual along $SO(3)/SO(2)$}};
\end{tikzpicture}
\]
\caption{Dualities for geometries with $S^3\times S^3$.}
\label{FigStmS}
\end{figure}
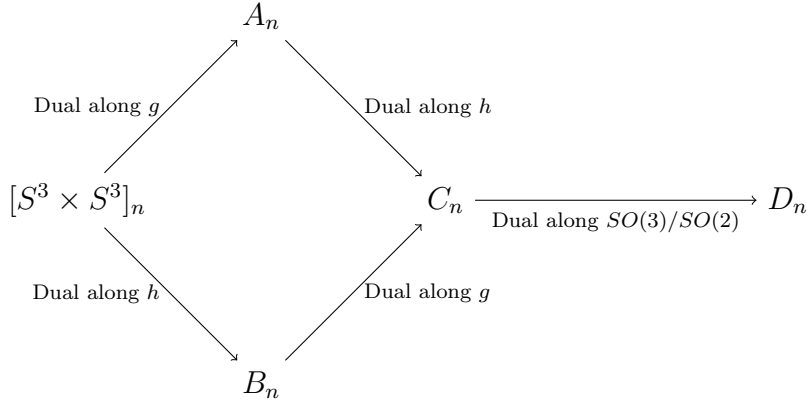
\begin{enumerate}
\item Starting with the pure metric (\ref{StartTsT}), one can perform a shift 
\bea
g\rightarrow {\tilde g}^n g,
\eea
which was called a spectral flow operation in \cite{LStSt} due to its interpretation in the dual field theory \cite{SchwSeib,SpecFlowW1,SpecFlowW2}. In the shifted coordinates, the metric becomes
\bea\label{SpecFlowR}
ds^2=F{\tilde \sigma}^a{\tilde \sigma}^a+G(R^a+{\tilde L}^{a(n)})(R^a+{\tilde L}^{a(n)})+G\sigma^a\sigma^a+ds_\perp^2
\eea
where
\bea
R^a=-i\mbox{Tr}(t^a dg g^{-1}),\quad 
{\tilde L}^{a(n)}=-i\mbox{Tr}\left[t^a {\tilde g}^{-n}d({\tilde g}^{n})\right]
\eea
We refer to \cite{LStSt} for the details.

In particular, a spectral flow by one unit ($n=1$) leads to the metric
\bea\label{SpecFlowRn1}
ds^2=F{\tilde \sigma}^a{\tilde \sigma}^a+G(R^a+{\tilde \sigma}^{a})(R^a+{\tilde \sigma}^{a})+G\sigma^a\sigma^a+ds_\perp^2\,,
\eea
which can also be rewritten as
\bea\label{SpecFlow}
ds^2=H({\tilde \sigma}^a+\frac{F}{H}D^{ae}\sigma^e)({\tilde \sigma}^a+\frac{F}{H}D^{ae}\sigma^e)+\frac{FG}{H}\sigma^a\sigma^a+ds_\perp^2
\eea
Here 
\bea
D^{ab}=\mbox{Tr}\left[t^agt^bg^{-1}\right],\quad H=F+G.
\eea
In contrast to (\ref{SignaGeneral}), the metric (\ref{SpecFlow}) contains information about the group elements $g$ not only through the left--invariant forms $\sigma_a$, but also through the matrix $D^{ab}$. 
\item Article \cite{LStSt} carried out various non--abelian T dualities for a large class of geometries (\ref{SpecFlowR}) supported by the Ramond--Ramond fluxes following the paths shown in figure \ref{FigStmS}. This procedure resulted in four non--abelian versions of the TsT transformations:
\bea
X_0\rightarrow X_n,\quad\mbox{where}\quad X=A,B,C,D.
\eea
As discussed in \cite{LStSt}, in the decompactification limit of the spheres, $B_n$ 
reduces to an abelian version of the TsT transformations, while $A_n$ becomes a trivial change of coordinates.

\item In contrast to the construction of \cite{LStSt}, which found non--abelian TsT transformations of the dual spaces $(A_0,B_0,C_0,D_0)$, in this article we will focus on the TsT deformations of the $S^3\times S^3$ itself. Specifically, we will look for a space $S_\gamma$ that satisfy three conditions:
\begin{enumerate}
\item The metric and the B field for $S_\gamma$ have the form (\ref{SignaGeneral}), where $f_k$ are functions of transverse coordinates.
\item When the deformation parameter $\gamma$ is set to zero, the space $S_0$ reduces to the geometry (\ref{StartTsT}) containing a product $S^3\times S^3$ with warp factors $(F,G)$ depending on the transverse coordinates.
\item In limit when spheres are decompactified, the space $S_\gamma$ reduces to the abelian version of the TsT transformation (\ref{TsTabel}). 
\end{enumerate}
Geometries with these three properties will be constructed in the next subsection, and their relation to the solutions found in \cite{LStSt} is summarized in figure \ref{FigS3S3paths}.

\end{enumerate}

\begin{figure}
\vskip -0.3cm
\[
\begin{tikzpicture}[scale=4.75]
\node (A) at (0,0.5) {$S^3\times S^3$};
\node (B) at (0.7,0.5) {$A_0$};
\node (F) at (0,0) {$[S^3\times S^3]_n$};
\node (D) at (0.7,0) {$A_n$};
\draw[->] (A) -- (F) node[midway,left]{\scriptsize{Shift}};
\draw[->] (A) -- (B) node[midway, above]{\scriptsize{T}};
\draw[->] (F) -- (D) node[midway, above]{\scriptsize{T}};
\node (Z) at (1.5,0.5) {$S^3\times S^3$};
\node (Y) at (2.2,0.5) {$A_0$};
\node (W) at (1.5,0) {$[S^3\times S^3]_\gamma$};
\node (X) at (2.2,0) {};
\draw[->] (Y) -- (X) node[midway,left]{\scriptsize{Shift}};
\draw[->] (Z) -- (Y) node[midway, above]{\scriptsize{T}};
\draw[->] (X) -- (W) node[midway, above]{\scriptsize{T}};
\node (S) at (0.4,-0.25) {(a)};
\node (T) at (1.9,-0.25) {(b)};
\end{tikzpicture}
\]
\caption{The duality paths corresponding to TsT deformations of $S^3\times S^3$ constructed in article \cite{LStSt} (a) and in this paper (b).}
  \label{FigS3S3paths}
\end{figure}
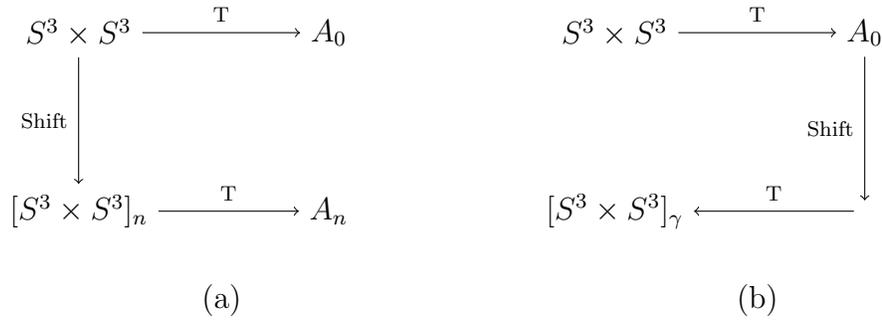

\subsection{Deformation of $S^3\times S^3$}
\label{SecSubS3S3}

In this subsection we will construct a non--abelian TsT deformation of the geometry (\ref{StartTsT}) containing a product of two three--spheres. As discussed above, the symmetries imply that the result has the form the  (\ref{SignaGeneral}). To be specific, we will focus on deformations of flat space, i.e., we will take  (\ref{StartTsT}) to be
\bea
ds^2=du^2+dv^2+u^2\sigma_a\sigma_a+v^2{\tilde\sigma}^a{\tilde\sigma}^a+
ds_{1,2}^2\,,
\eea
where $ds_{1,2}^2$ describes a three--dimensional Minkowski space. To make some symmetries explicit, we slightly modify the notation for the string metric (\ref{SignaGeneral}) of the deformed space and write the final geometry as
\bea\label{SignaGeneralFlat}
ds_S^2&=&du^2+dv^2+\frac{(u f_1)^2}{4}\sigma_a\sigma_a+
\frac{(vf_2)^2}{4}{\tilde\sigma}^a{\tilde\sigma}^a+
2 \frac{uvf_3}{4}{\sigma}^a{\tilde\sigma}^a+ds_{1,2}^2\nn
B&=&\gamma f_4 (uv)^2\sigma_a\wedge {\tilde\sigma}_a,\quad e^{-2\Phi}=(f_5)^3
\eea
We are interested in solutions which reduce to (\ref{TsTabel}) in the decompactification limit. Specifically, making replacements 
\bea
\sigma_a\rightarrow \eps dx_a,\quad {\tilde\sigma}_a\rightarrow \eps dy_a,\quad 
u\rightarrow \frac{F}{\eps}+{\bar u},\quad v\rightarrow \frac{G}{\eps}+{\bar v}
\eea
and expanding the geometry to the {\it first order} in $\eps$, we expect to recover the solution with 
\bea
f_1=f_2=\left[1+\gamma^2 FG\right]^{-\frac{1}{2}},\quad f_3=0,\quad 
f_4=\frac{1}{1+\gamma^2 FG},\quad f_5=1+\gamma^2 FG
\eea
Motivated by this result, we expand various functions in power series as
\bea\label{PertElltst}
f_\ell=\sum P^{(\ell)}_k(u,v) \gamma^{2k},\quad \ell=1,\dots,5.
\eea
and solve for functions $P^{(\ell)}_k(u,v)$. Assuming that $\gamma$ is the only new scale in the problem\footnote{Some solutions violating this assumption will be discussed in the next subsection, but they don't describe the TsT deformations.}, dimensional analysis implies that $P^{(\ell)}_k(u,v)$ is a homogeneous polynomial of degree $(4k)$ in $u$ and $v$, and equations of motion uniquely fix the expansions (\ref{PertElltst}).

Although the ansatz (\ref{SignaGeneralFlat}) is written in the string frame, equations of motion take the simplest form for the rescaled metric $g_{\mu\nu}^{(E)}$ defined by
\bea
g_{\mu\nu}^{(E)}=e^{\psi} g_{\mu\nu}^{(s)},\quad \psi=-\frac{4\Phi}{d-2}
\eea
Then the effective action for the NS--NS fields
\bea
S&=&\int d^d x\sqrt{-g}\left[R-\frac{4}{d-2}(\d\Phi)^2-\frac{1}{12}e^{2\psi} 
H_{\mu\nu\sigma}H^{\mu\nu\sigma}\right]
\eea
leads to equations of motion
\bea
R_{\mu\nu}-\frac{1}{2}g_{\mu\nu}R&=&
\frac{d-2}{4}\left[\d_\mu\psi\d_\mu\psi-\frac{1}{2}g_{\mu\nu}(\d\psi)^2\right]+\frac{e^{2\psi}}{4}
\left[H_{\mu \alpha\beta}{H_\nu}^{\alpha\beta}-\frac{1}{6}g_{\mu\nu}H^2\right]\nn
\nabla^2\psi&=&\frac{4}{12(d-2)}e^{2\psi} 
H_{\mu\nu\sigma}H^{\mu\nu\sigma},\quad 
\nabla_\mu\left[e^{2\psi} H^{\mu\nu\sigma}\right]=0
\eea
Substituting the ansatz (\ref{SignaGeneralFlat}) with expansions (\ref{PertElltst}) into these equations, we find the {\it unique} solution order--by--order in $\gamma$:
\bea\label{TsTsolnS3}
&&f_1=f_2=1-16(\gamma u v)^2+\mathcal{O}(\gamma^4),\quad 
f_3=\frac{16\gamma^2}{3}(u v)(u^2+v^2)+\mathcal{O}(\gamma^4),\nn
&&f_4=\left[1+\gamma^2\left(\frac{16(uv)^2}{9}+c\left[3u^4-8(uv)^2+3v^4\right]\right)\right]+\mathcal{O}(\gamma^4),\\
&&f_5=1+{16(\gamma u v)^2}+\mathcal{O}(\gamma^4)\nonumber
\eea
To fix the value of $c$, one has to expand equations to order $\gamma^8$. It is also interesting to look at the inverse matrix of the combination $(g^{(s)}+B)$ which plays an important role in the abelian T duality\footnote{Comparing this result with the general classification of integrable deformations of spheres \cite{BDH21}, we conclude that (\ref{GBinvS3S3}) viewed as a deformation of the sigma model does not preserve integrability.}:
\bea\label{GBinvS3S3}
&&\hskip -1.4cm(g^{(s)}+B)^{\mu\nu}\d_\mu \d_\nu=\d_u^2+\d_v^2\\
&&\hskip -1cm+4\sum_a
[\tau_a\,\,{\tilde\tau}_a]
\left[\begin{array}{cc}
u^{-2}&-4\gamma\left[\frac{4\gamma^2}{3}(u^2+v^2)+1\right] \\
-4\gamma\left[\frac{4\gamma^2}{3}(u^2+v^2)-1\right]&v^{-2}
\end{array}
\right]
\left[\begin{array}{c}
\tau_a \\ {\tilde\tau}_a
\end{array}
\right]+\mathcal{O}(\gamma^3)\nonumber
\eea
Here $(\tau_a,{\tilde\tau}_a)$ are inverted left--invariant forms defined by
\bea
\tau_a=\tau_a^\mu\d_\mu,\quad {\tilde\tau}_a={\tilde\tau}_a^\mu\d_\mu,\quad
\sigma^a_\mu \tau_b^\mu=\delta_b^a,\quad {\tilde\sigma}^a_\mu {\tilde\tau}_b^\mu=\delta_b^a
\eea
Decompactification limit is obtained by writing
\bea
u=\frac{c_1}{\eps}+{\bar u},\quad v=\frac{c_2}{\eps}+{\bar v}, \quad
(\sigma,{\tilde\sigma})={\eps}(dx,dy),\quad\gamma=\eps^2{\bar\gamma}
\eea
and expanding (\ref{GBinvS3S3}) up to the linear order in $\eps$. As expected, this recovers the result for the abelian TsT transformation applied to $T^3\times T^3$.

\subsection{Deformation of $S^3\times T^3$}
\label{SecSubS3T3}

\begin{figure}
\begin{tabular}{ccc}
\hskip -0.6cm \ 
  \includegraphics[width=0.5 \textwidth]{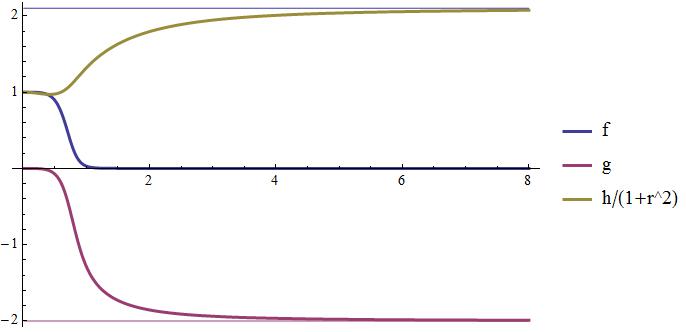}&\ \ \ \ \  &
  \hskip -0.8cm \ 
   \includegraphics[width=0.47 \textwidth]{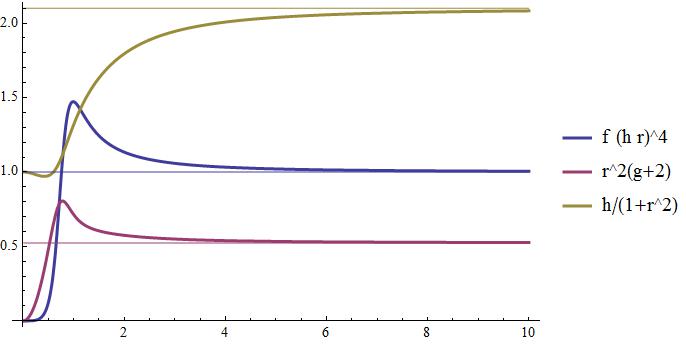}\\
(a)&&(b)
\end{tabular}
\caption{Profiles of various functions appearing in the ansatz (\ref{SigmaGenYnew}) for the gravitational TsT deformation of $S^3\times T^3$: (a) general picture; (b) behavior at infinity and numerical extraction of parameters $(\alpha,\beta)$ for $\gamma=1$.
}
\label{FigTsTtor}
\end{figure}

In this subsection we consider the limit of the geometry (\ref{SignaGeneral}) when one of the spheres is decompactified. Making a replacement ${\tilde\sigma}_b\rightarrow dy_b$ in (\ref{SignaGeneral}), one finds the expressions for all NS--NS fields in the string frame:
\bea\label{SigmaGenYbFrm}
ds^2=f_1 \sigma_a\sigma_a+f_2 dy_bdy_b+
f_3\sigma_a dy_a+ds_\perp^2,
\quad B=f_4 \sigma_a\wedge dy_a,\quad e^{2\Phi}=f_5
\eea
Comparing this with (\ref{TsTabel}), we conclude that a candidate for the TsT transformation should have certain symmetry properties under $y_a\rightarrow -y_a$: the metric should remain invariant, while the $B$ field should flip sign. This implies that the TsT deformation must have $f_3=0$. Starting with such solution and dualizing along $y$ directions, we arrive at a pure metric:
\bea\label{SigmaGenY}
ds^2=f_1 \sigma_a\sigma_a+\frac{1}{f_2}(dy_a+f_4 \sigma_a)(dy_a+f_4 \sigma_a)+
ds_\perp^2,
\quad B=0,\quad e^{2\Phi}=1.
\eea
As in section \ref{SecSubS3S3}, we focus on deformation of the flat space with the metric 
\bea
ds^2=dr^2+r^2 \sigma_a\sigma_a+dy_ady_a+ ds_{2,1}^2.
\eea
Rather than keeping the transverse space undeformed as in (\ref{SigmaGenY}), it is convenient to make a change of coordinates so that it is the $g_{ab}$ components of the metric that remain fixed and to parameterize the geometry (\ref{SigmaGenY}) as\footnote{We also rearranged several terms and relabeled various functions in (\ref{SigmaGenY}) to simplify the resulting Einstein's equations. We also dropped irrelevant transverse coordinates.}
\bea\label{SigmaGenYnew}
ds^2=f dr^2+\frac{r^2}{4} (\sigma_a+hG dy_a) (\sigma_a+hG dy_a)+\frac{G^2}{4} dy_a dy_a\,.
\eea
The TsT deformation of $S^3\times T^3$ is obtained from (\ref{SigmaGenYnew}) by performing T dualities along the torus:
\bea\label{SigmaGenYtSt}
ds^2&=&f dr^2+\frac{(rG)^2}{16}e^{2\Phi} \sigma_a\sigma_a+e^{-2\Phi} dy_a dy_a,\\
B&=&e^{2\Phi}\frac{r^2hG}{4} dy_a\wedge \sigma_a,\quad e^{2\Phi}=\frac{4}{G^2[1+(rh)^2]},\nonumber
\eea
The ansatz (\ref{SigmaGenYnew}) has a residual symmetry under a simultaneous rescaling of coordinates $y_a$ and function $G$, so the order of differential equations can be lowered by introducing a new function $g$ through a relation
\bea
G=\exp\left[\int \frac{g}{r} dr\right]\,.
\eea
The the rescaling symmetry implies that all integrals disappear from the Einstein's equations, which reduce to 
\bea\label{EinstS3T3simple}
&&2r {\dot g}+4 f g+{4 f (g+2) h^4 r^4+(g+1)\left[ (rgh)^2+2 g h {\dot h} r^3+{\dot h}^2 r^4\right]}=0,\nn
&&(1-r^4 h^4)f-(1+3g+g^2)+\frac{r^2}{4}\left[{g h+r{\dot h}}\right]^2=0,
\eea
and one more rather cumbersome equation (\ref{EinstS3T3compl}) which disappears if $h=0$. All three equations are invariant under simultaneous rescaling of $r$ and $h$ with a fixed product $rh$. Motivated by the abelian TsT transformation, we assume that the only scale in the problem is given by a single deformation parameter $\gamma$. This implies that function $(f,G,h/r)$ must depend on a single scale--invariant combination $x=\gamma r$, then equations (\ref{EinstS3T3simple}), (\ref{EinstS3T3compl}) lead to the {\it unique} series expansions for all ingredients:
\bea\label{TsTexpand}
f&=&1-x^4-\frac{19 x^6}{9}-\frac{23 x^8}{9}-\frac{55 x^{10}}{54}+\mathcal{O}(x^{12}),\nn
G&=&1-\frac{x^4}{6}-\frac{13 x^6}{54}-\frac{17 x^8}{72}
-\frac{101 x^{10}}{810}+\mathcal{O}(x^{12}),\\
Gh&=&\gamma\left[1+\frac{2 x^2}{3}+\frac{x^4}{2}+\frac{7 x^6}{27}-\frac{7 x^8}{81}+\mathcal{O}(x^{10})\right].\nonumber
\eea
By continuing these expansions, one arrives at the unique deformation (\ref{SigmaGenYnew}), but unfortunately we were not able to extract the analytical expressions for various functions in closed form. Numerical integration of the system (\ref{EinstS3T3simple}), (\ref{EinstS3T3compl}) leads to profiles for various functions shown in figure \ref{FigTsTtor}(a). 

Alternatively, one may expand solutions of the system (\ref{EinstS3T3simple}), (\ref{EinstS3T3compl}) near infinity, and the result is fully specified by two free parameters $(\alpha,\beta)$:
\bea
g&=&-2+\frac{\alpha}{r^2}+\frac{\alpha^2}{r^4}+\frac{\alpha^3}{2r^6}-\frac{2}{\beta^2 r^6}+
\mathcal{O}(r^{-8}),\nn
h&=&\beta\left[r^2+\frac{\alpha}{2}+\frac{3\alpha^2}{8r^2}+\frac{11\alpha^3}{48r^4}-\frac{1}{\beta^2 r^4}+\mathcal{O}(r^{-6})\right]\\
f&=&\frac{1}{\beta^4 r^{12}}\left[1-\frac{\alpha}{r^2}-\frac{\alpha^2}{r^4}-\frac{\alpha^3}{6r^6}+\frac{6}{\beta^2 r^6}+\mathcal{O}(r^{-8})\right]\nonumber
\eea
These parameters are uniquely fixed in terms of $\gamma$ by integrating through the intermediate region, e.g., by using numerics. Figure \ref{FigTsTtor}(b) shows an example of extracting these parameters for $\gamma=1$: 
\bea
\gamma=1\quad \Rightarrow\quad \alpha=0.524,\quad \beta=2.10
\eea
We conclude this subsection by making some comments about several gravity solutions closely related to (\ref{SigmaGenYnew}). 
\begin{enumerate}[(a)]
\item Geometries (\ref{SigmaGenYnew}) and (\ref{SigmaGenYtSt}) describe two families of one--parameter deformations of $S^3\times T^3$ which have either off-diagonal terms in the metric or a non--trivial $B$ field. It is interesting to construct a more general deformation, where both effects are turned on, and we do so in the Appendix \ref{AppS3T3extra}. The final results are given by equations (\ref{S3S3TwoPar2})--(\ref{S3S3TwoPar2a}). In contrast to the abelian  (i.e., $T^3\times T^3$) case, where the off--diagonal terms in the metric can be removed by a coordinate transformation, for $S^3\times T^3$ both deformation parameters lead to new geometries.

\item Focusing on diagonal solutions with one deformation, i.e., setting $h=0$ in (\ref{SigmaGenYnew}), we arrive at a drastic simplification in 
(\ref{EinstS3T3simple}),
\bea
f=1+3g+g^2,\quad 
r{\dot g}+2g(1+3g+g^2)=0\,,
\eea
and the last remaining equation (\ref{EinstS3T3compl}) disappears. Equation for $g$ can be easily integrated,
\bea\label{Spurious}
\frac{r}{r_0}=\frac{5+3\sqrt{5}}{20}\ln\left[2+\frac{3-\sqrt{5}}{g}\right]+
\frac{5-3\sqrt{5}}{20}\ln\left[2+\frac{3+\sqrt{5}}{g}\right]
\eea
but the resulting geometries (\ref{SigmaGenYnew}) and (\ref{SigmaGenYtSt}) have naked singularities. In contrast to the expansions (\ref{TsTexpand}), which give the unique solution with one one scale $\gamma$, solution (\ref{Spurious}) has an additional scale $r_0$. By putting back the off--diagonal terms, one can construct geometries (\ref{SigmaGenYnew}) and (\ref{SigmaGenYtSt}) with two parameters $(\gamma,r_0)$, but the resulting cumbersome expressions do not have a clear physical interpretation, in contrast to (\ref{TsTexpand}).
\item A decompactification limit of the three--sphere in (\ref{SigmaGenYnew}) inspires an ansatz
\bea\label{AnstzT3T3}
ds^2=dr^2+{F}^2\left[dx_a+hG dy_a \right]^2+{G}^2 (dy_a)^2
\eea
where  $(F,G,h)$ are functions or $r$ only. Expanding these functions in Taylor series,  one observes that $h$ can be removed by a linear transformation of $(x,y)$ with constant coefficients, so we can set $h=0$ without loss of generality. The symmetry under simultaneous rescalings of 
$(F,x)$ or $(G,y)$ suggest the change of variables from $(F,G)$ to $(A,B)$ defined by
\bea
F=\exp\left[\int A dr\right],\quad G=\exp\left[\int B dr\right]\,.
\eea
Integration constants can be absorbed into rescalings of $(x,y)$. The Einstein's equations reduce to three relations,
\bea
A^2+3 AB+B^2=0,\quad
{\dot A}=-3A(A+B),\quad {\dot B}=-3B(A+B)\,,
\eea
which can be integrated to give
\bea
A=\frac{1+\mu}{6(r+r_0)},\quad B=\frac{1-\mu}{6(r+r_0)}\quad \mu=\pm \sqrt{5}\,.
\eea
Choosing the plus sign in the last relation, we arrive at the most general solution of the Einstein's equations with the ansatz (\ref{AnstzT3T3}):
\bea\label{toriSoln}
F=F_0\left[1+\frac{r}{r_0}\right]^{\frac{1+\sqrt{5}}{6}},\quad
G=G_0\left[1+\frac{r}{r_0}\right]^{\frac{1-\sqrt{5}}{6}},\quad h=\frac{h_0}{G}
\eea
Here $(F_0,G_0,h_0,r_0)$ are arbitrary constants. In the Appendix \ref{AppTor} we discuss the recovery of this solution in the decompactification limit of (\ref{toriSoln}) and present a generalization of (\ref{toriSoln}) to a product of tori of arbitrary dimensions. Note that the decompactification limit of (\ref{TsTexpand}), which has only one scale $\gamma$, is recovered by sending $r_0$ to infinity in (\ref{toriSoln}) and setting $h_0=\gamma G_0$. In this limit, the geometry (\ref{AnstzT3T3}) describes a constant twist between two tori.
\end{enumerate}

\subsection{Deformation of $S^n\times T^n$ for $n\ne 3$}
\label{SecSubSnTn}

In this and next subsections we will extend the proposed ``gravitational TsT deformation'' to geometries containing products of two $n$--dimensional spheres with $n\ne 3$. We will see that all these cases can be treated uniformly, and the resulting construction is rather different from the one for $S^3\times S^3$. We will begin with a detailed discussion of the deformed $S^n\times T^n$ space and conclude with presenting results for $S^n\times S^n$ which are obtained using the same logic, but which is more technically involved.

Let us consider the $2n+1$--dimensional flat space written in terms of an $n$--dimensional sphere and an $n$--torus:
\bea\label{TnSnStart}
ds^2=dr^2+r^2d\Omega_n^2+ds_{T^n}^2=
dr^2+r^2\frac{4dx_adx_a}{[1+\rho^2]^2}+dy_bdy_b\,,\quad \rho^2=x_ax_a
\eea
This parameterization has an explicit $SO(n)\times SO(n)$ symmetry under independent rotations of coordinates $x_a$ and $y_b$.
To extend the results obtained in section \ref{SecSubS3T3}, we consider a deformation of this metric that preserves the diagonal $SO(n)$ as well as translational invariance along the torus directions. The result must be written in terms of covariant objects,
\bea\label{Aug15}
dx_a,\quad dy_a,\quad x_a,\quad \delta_{ab},\quad \eps_{a_1\dots a_n}
\eea
and all indices must be contracted in the end. Since a product of two Levi-Civita symbols can be reduced to products of Kronecker deltas, the metric can be rewritten in the form where 
$\eps_{a_1\dots a_n}$ enters at most once in every term. At this point we encounter a drastic difference between the $n\ge 4$ and lower--dimensional cases: while it is possible to have some terms with Levi-Civita symbols in the latter cases, this cannot happen for $n>3$. Indeed, since the Levi-Civita symbol has to contract with the first three objects in (\ref{Aug15}), and the metric is quadratic in differentials, one necessarily encounters a contraction $[x_a x_b\eps_{ab\dots}\dots]$ that vanishes. Therefore, for $n\ge 4$, the most general deformation of the metric (\ref{TnSnStart}) preserving the diagonal $SO(n)$ and translational invariance in the torus directions is 
\bea\label{TnSnDef}
ds^2&=&dr^2+r^2 h_1{dx_adx_a}+h_2 dy_bdy_b+r^2 h_3(x_a dx_a)^2+
r h_4(x_a dy_a)^2\nn
&&+2\gamma h_5(x_a dx_a)(x_b dy_b)
+2\gamma h_6 dx_ady_a,
\eea
where $(h_1,\dots,h_6)$ are functions of $(r,\rho)$ and the deformation parameter $\gamma$. Furthermore, motivated by the abelian TsT transformation, we will impose a $\mathbb{Z}_2$ symmetry that flips the signs of $\gamma$ and torus directions. Then $(h_1,\dots h_6)$ must be even functions of $\gamma$. 

Before analyzing the metric (\ref{TnSnDef}), let us make some comments about $n=2,3$. In section \ref{SecSubS3T3} we studied deformations of $S^3\times T^3$ and wrote the metric of the three--dimensional sphere in terms of the left--invariant forms. Alternatively, this metric can be written in terms of coordinates $x_a$ as in (\ref{TnSnStart}), and the map between these coordinates and the left invariant forms was found in \cite{LSdft}
\bea\label{SigmaAsX}
\sigma_a=\frac{1}{[1+\rho^2]^2}\left[(1-\rho^2)dx_a+2x_a x_b dx_b-2\eps_{abc}x_bdx_c\right].
\eea
By substituting these expressions in (\ref{SigmaGenY}) and rearranging terms, we would arrive at the metric (\ref{TnSnDef}) with several additional terms involving the Levi-Civita symbols. Since all coefficients 
in (\ref{SigmaAsX}) are fixed by 
the $SU(2)$ invariance, the metric (\ref{TnSnDef}) with additional terms is fully determined by $(h_1,h_2,h_6)$, as expected from (\ref{SigmaGenY}). In the case of $S^2$, one can add several terms with $\eps_{ab}$ to (\ref{TnSnDef}), and the most general metric is given by (\ref{S2T2anstz}). The consequences of these additional terms are analyzed in the Appendix \ref{AppS2T2}, where it is shown that the answers reduce to those with $n\ge 4$ up to an $SO(2)$ rotation of coordinates $y_a$. In this section we will focus on analyzing the metric (\ref{TnSnDef}) for $n>3$. 

\bigskip 

Although conceptually deformations of all $S^n\times T^n$ with $n>3$ work in the same way, some minor technical details are different, so to avoid unnecessary clutter we will focus on $n=4$ and comment on general $n$ in the end of this subsection. 

\bigskip

Let us begin with analyzing the Einstein's equations for the metric (\ref{TnSnDef}) in the linear order in $\gamma$. In this approximation, functions $(h_1,\dots, h_4)$ should be taken from (\ref{TnSnStart}), and $(h_5,h_6)$ must satisfy three Einstein's equations
\bea\label{SnTnEinst}
&&2(1+y)\d_y\d_r h_6-4\d_r h_6+2y(1+y)\d_r\d_y h_5+(y+5)\d_r h_5=0,\nn
&&(y+1)^2(\d_y h_5-2\d_y^2 h_6)+2r(2\d_r h_5+r\d_r h_5)=0,\\
&&-3(y+1)^2(h_5-2\d_y h_6)+4r\left[2\d_r (h_6+yh_5)+r\d^2_r(h_6+y h_5)\right]=0\,.\nonumber
\eea
The most general solution of these equations is
\bea\label{SnTnSoln}
h_5=\frac{4g-2(1+y)\d_y g-{\tilde h}(y)}{3(y+1)},\quad 
h_6=g-y h_5\,,
\eea
where ${\tilde h}(y)$ is an arbitrary function, and $g$ satisfies a PDE:
\bea\label{SnTnSoln1}
\d_r[r^2\d_r g]+\frac{(1+y)^4}{y^2}\d_y\left[\frac{y^3\d_y g}{(1+y^2)^2}\right]-
(y+3)g+\frac{(1+y)^2}{2\sqrt{y}}\d_y\left[\frac{y^{3/2}{\tilde h}}{1+y^2}\right]=0.
\eea
Equations (\ref{SnTnSoln})--(\ref{SnTnSoln1}) describe deformations of the flat space with an infinite numbers of parameters. The TsT transformation is expected to correspond to a deformation with only one parameter that has a specific dimension\footnote{In particular, this parameter is dimensionless for the standard abelian TsT.}, then dimensional analysis implies that
\bea\label{h5h6Separ}
h_5=r^k {\bar h}_5(y),\quad h_6=r^k {\bar h}_6(y)
\eea
Note that this assumption is consistent with the structure of equations (\ref{SnTnSoln})--(\ref{SnTnSoln1}): since they are invariant under rescaling of $r$, all separable solutions have the form (\ref{h5h6Separ}). 
Substitution of this ansatz into the Einstein's equations (\ref{SnTnEinst}) gives
\bea\label{SnTnKsoln}
k=0:&&{\tilde h}_5=2\d_y {\tilde h}_6,\quad {\mbox{arbitrary}}\quad {\tilde h}_6\,;\nn
k=1:&&{\tilde h}_5=\frac{y^2+2y+9}{y+1},\quad {\tilde h}_6=\frac{2(y+5)}{y+1}\,;\\
k\ge 2:&&{\tilde h}_5=\frac{4k(k+1){\tilde h}_6+6(1+y)^2\d_y {\tilde h}_6}
{3y^2-(2k+1)^2y+7y+3},\quad
{\tilde h}_6=\frac{P_k[y]}{(y+1)^{k+1}}\,.\nonumber
\eea
Here $P_k[y]$ are specific polynomials of degree $k$ with coefficients which are not illuminating. 
Equation for ${\tilde h}_6$ also has the second solution which diverges when $y$ goes to zero, and therefore does not have a good decompactification limit, where one writes
\bea\label{DecompT4S4}
r=\frac{1}{\eps}+{\bar r},\quad x_a\rightarrow \eps {\bar x}_a
\eea
and sends $\eps$ to zero while keeping $({\bar r},{\bar x}_a)$ fixed\footnote{As we will discuss below, the decompactification limit may also involve a rescaling of the deformation parameter, but this does not affect the argument of regularity.}. Therefore, we will focus on exploring the regular options (\ref{SnTnKsoln}).

\bigskip
\noindent
To understand the properties of the solutions (\ref{SnTnKsoln}) with various values of $k$, it is instructive to start with analyzing some specific cases:
\begin{enumerate}[(a)]
\item For $k=0$, the second line of the metric (\ref{TnSnDef}) becomes
\bea
ds_\gamma^2&=&\frac{\gamma}{\rho} \d_\rho {\tilde h}_6(x_a dx_a)(x_b dy_b)
+2\gamma {\tilde h}_6 dx_ady_a=2\gamma d({\tilde h}_6 x_a)dy_a
\eea
This describes a diffeomorphism $y_a\rightarrow y_a+\gamma {\tilde h}_6 x_a$ of the undeformed metric. In the decompactification limit, when the translational invariance in $x$ coordinates is recovered, this transformation reduces to a shift 
$y_a\rightarrow y_a+\gamma x_a$.

The diffeomorphism  $y_a\rightarrow y_a+\gamma {\tilde h}_6(\rho) x_a$ can be applied even beyond the linear order in $\gamma$ leading to the {\it exact} metric that has the form (\ref{TnSnDef}). 
\bea
ds^2&=&dr^2+{dx_adx_a}+(dy_b+\gamma d[{\tilde h}_6(\rho) x_b])^2\nonumber
\eea
Dualization along the torus leads to a metric of flat space and a Kalb--Ramond field with vanishing field strength, therefore, the $k=0$ option does not lead to interesting deformations.
\item For $k=1$,  the second line of the metric (\ref{TnSnDef}) becomes
\bea\label{TsTS4pertK1}
ds_\gamma^2&=&\frac{2\gamma r}{(1+\rho^2)}\left[2(\rho^2+5)(x_a dx_a)(x_b dy_b)
+(9+2\rho^2+\rho^4)dx_ady_a\right]
\eea
This deformation is singular at $r=0$, as can be seen by going to Cartesian coordinates in a vicinity of this point. Therefore, this is not a good candidate for a non--abelian TsT transformation. 

Furthermore, perturbation (\ref{TsTS4pertK1}) does not have a natural interpretation in the decompactification limit, where one performs a change of variables (\ref{DecompT4S4}) 
and expands the metric to the {\it first order} in $\eps$ while keeping $({\bar r},{\bar x}_a)$ fixed. 
\item For $k=2$,  the second line of the metric (\ref{TnSnDef}) becomes
\bea\label{TsTS4pert}
ds_\gamma^2&=&\frac{2\gamma r^2}{(1+\rho^2)^2}\left[2(x_a dx_a)(x_b dy_b)
+(1-\rho^2)dx_ady_a\right]
\eea
In the decompactification limit (\ref{DecompT4S4}), this correction describes a shift 
\bea
x_a\rightarrow x_a+\gamma y_a\quad \Rightarrow \quad
{\bar x}_a\rightarrow {\bar x}_a+\frac{\gamma}{\eps} y_a\,,
\eea
which gives rise to the abelian TsT transformation upon dualization along $T^4$. The T dual of the perturbation (\ref{TsTS4pert}) is the most natural candidate for an infinitesimal non--abelian TsT transformation. 
\item For $k=3$,  the second line of the metric (\ref{TnSnDef}) becomes
\bea
ds_\gamma^2&=&\frac{2\gamma r^3}{(1+\rho^2)^3}\left[(10-6\rho^2)(x_a dx_a)(x_b dy_b)
+(1-3\rho^2)(3-\rho^2)dx_ady_a\right]
\eea
The decompactification limit (\ref{DecompT4S4}) gives
\bea\label{TsTS4pertK3}
ds_\gamma^2&\simeq&\frac{6\gamma}{\eps^2} (1+3\eps {\bar r})d{\bar x}_ady_a\simeq
\frac{9\gamma}{\eps^2} (1+2\eps {\bar r})d{\bar x}_ady_a-
\frac{3\gamma}{\eps^2}d{\bar x}_ady_a
\eea
These terms can be interpreted as simultaneous  shifts
\bea
{\bar x}_a\rightarrow {\bar x}_a+\frac{9\gamma}{2\eps^2}y_a,\quad 
y_a\rightarrow y_a-\frac{3\gamma}{2\eps^2}{\bar x}_a,
\eea
which remain finite if $\gamma$ is scaled as $\eps^2$. Therefore, in the decompactification limit, the perturbation (\ref{TsTS4pertK3}) reduces to a combination of options (a) and (c) with an appropriate scaling of $\gamma$. Similarly, decompactification limits of perturbations with $k>3$ reduces to combinations of options (a) and (c) as long as $\gamma$ is scaled as 
$\eps^{k-1}$.

We conclude that the {\it unique} TsT transformation in the decompactification limit gives rise to an infinite discrete family of transformations once such limit is removed. We will focus on the most interesting case of $k=2$. 
\end{enumerate}
\noindent
To summarize, we will study the deformation that has the form (\ref{TsTS4pert}) in the linear order in $\gamma$. This implies that $\gamma$ has the dimension of inverse length, so on dimensional grounds, various functions appearing in (\ref{TnSnDef}) must have expansions 
\bea
h_k=\sum_{p=0}^\infty (\gamma r)^p h_k^{(p)}(\rho),\ k=1\dots 4;\qquad
h_{5,6}=r^2 \sum_{p=0}^\infty (\gamma r)^p h_{5,6}^{(p)}(\rho)
\eea
The leading terms in the expansions of $h_{5,6}$ are summarized in (\ref{TsTS4pert}). Solving Einstein's equations to the second order in $\gamma$, we find 
\bea\label{AnswS4T4}
ds^2&=&dr^2+r^2\frac{4dx_adx_a}{[1+\rho^2]^2}+\left[1+\frac{(\gamma r)^2}{4}-
\frac{(3+c)\eta (\gamma r\rho)^2}{4(1+\rho^2)^2}
\right]dy_bdy_b+\frac{c(\gamma r)^2}{(1+\rho^2)^2}(x_a dy_a)^2\nn
&&+\frac{2\gamma r^2}{(1+\rho^2)^2}\left[2(x_a dx_a)(x_b dy_b)
+(1-\rho^2)dx_ady_a\right]+\mathcal{O}(\gamma^3)\,.
\eea
Here $c$ is a free parameter and $\eta=1$. In the decompactification limit,
\bea\label{DecompLim}
r=\frac{1}{\eps}+{\bar r},\quad x_a\rightarrow \eps {\bar x}_a,\quad  
\gamma\rightarrow \eps{\bar \gamma}\,,
\eea
we find
\bea
ds^2&=&d{\bar r}^2+(1+\eps{\bar r})^2\left[2d{\bar x}_a+\frac{\bar\gamma}{2}dy_a\right]^2+dy_bdy_b
+\mathcal{O}(\eps^2)+\mathcal{O}(\gamma^3)\,.
\eea
As expected, this geometry describes shifts of ${\bar x}_a$ coordinates by $y_a$, and it solves the Einstein's equations without $\mathcal{O}(\gamma^3)$ corrections. 

After dualizing (\ref{AnswS4T4}) along $T^4$, we find the geometry that solves equations of motion up to 
$\mathcal{O}(\gamma^3)$ terms\footnote{We have set $\eta=1$.}: 
\bea\label{SnTnTsT}
ds^2&=&dr^2+r^2\frac{4dx_adx_a}{[1+\rho^2]^2}+\left[1-\frac{(\gamma r)^2}{4}+
\frac{(3+c)(\gamma r\rho)^2}{4(1+\rho^2)^2}
\right]dy_bdy_b\nn
&&-\frac{c(\gamma r)^2}{(1+\rho^2)^2}(x_a dy_a)^2
-\frac{\gamma^2 r^4}{(1+\rho^2)^4}\left[2x_a(x_b dx_b)
+(1-\rho^2)dx_a\right]^2\,,\\
e^{-2\Phi}&=&1+{(\gamma r)^2}{}-
\frac{3(\gamma r\rho)^2}{(1+\rho^2)^2},\quad
B=\frac{\gamma r^2 e^{2\Phi}}{(1+\rho^2)^2}dy_a\wedge \left[x_a d\rho^2
+(1-\rho^2)dx_a\right]\,.\nonumber
\eea
In the decompactification limit, this geometry describes the TsT transformation of $T^4\times T^4$. 

Let us conclude this subsection with a comment on deformations of $S^n\times T^n$ geometries for arbitrary $n$. The $n=3$ case is rather special, and it  was discussed in section \ref{SecSubS3T3}. All $n\ge 4$ appear on the same footing, and direct calculations show that the metric (\ref{AnswS4T4}) gives the unique solution for all of them, but the value of the parameter $\eta$ depends in the dimensions of the sphere as
\bea\label{SnTnEta}
\eta_n=\frac{4(c+n-1)}{n(c+3)}\,.
\eea
For $n=2$, one has additional contributions to the ansatz (\ref{TnSnDef}) consistent with expected symmetries, and the consequences of these additions are analyzed in Appendix \ref{AppS2T2}. It turns out that the extra terms can be eliminated by rotating coordinates $(y_1,y_2)$, and the final answer (\ref{S2T2final}) matches the form (\ref{TnSnDef}). However, in contrast to (\ref{AnswS4T4}), which gives the most general solution for all $n\ge 2$, the result (\ref{S2T2final}) has three rather than two parameters. Once $c_1$ is set to zero, one recovers (\ref{AnswS4T4}) with the value of $\eta$ given by (\ref{SnTnEta}). The TsT deformation of $S^n\times T^n$  with $n\ne 3$ is given by (\ref{SnTnTsT}), the dual of (\ref{AnswS4T4}).

\subsection{Deformation of $S^n\times S^n$ for $n\ne 3$}
\label{SecSubSnSn}

In this short subsection we will analyze the deformations of 
$S^n\times S^n$. Rather than attempting to find the most general solutions of partial differential equations (all functions depend on two coordinates, the radii of undeformed spheres), we will use an inspiration from the results of the previous subsection to propose a very restrictive ansatz and verify that all equations of motion are satisfied up to 
$\mathcal{O}(\gamma^2)$.

\bigskip

Solution (\ref{AnswS4T4}) inspires a metric deformation of $S^n\times S^n$:
\bea\label{S4S4metr}
\hskip -0.5cm
ds^2=du^2+dv^2+u^2\frac{4dx_adx_a}{[1+\rho^2]^2}+v^2\frac{4dy_ady_a}{[1+\sigma^2]^2}
+\frac{2\gamma (uv)^2}{[(1+\rho^2)(1+\sigma^2)]^2}X_aY_a+\mathcal{O}(\gamma^2)\,,
\eea
where 
\bea
X_a=2x_a (x_b dx_b)+(1-\rho^2)dx_a,\quad Y_a=2y_a (y_b dy_b)
+(1-\sigma^2)dy_a
\eea
and
\bea
\rho^2=(x_a)^2,\quad \sigma^2=(y_a)^2.\nonumber
\eea
Direct calculations show that (\ref{S4S4metr}) satisfies Einstein's equations. 

Similarly, solution (\ref{SnTnTsT}) inspires a TsT--type deformation of $S^n\times S^n$:
\bea\label{S4S4bFrame}
ds^2&=&du^2+dv^2+u^2\frac{4dx_adx_a}{[1+\rho^2]^2}+v^2\frac{4dy_ady_a}{[1+\sigma^2]^2}+\mathcal{O}(\gamma^2).\nn
B&=&\frac{2\gamma (uv)^2}{[(1+\rho^2)(1+\sigma^2)]^2}X_a\wedge Y_a+
\mathcal{O}(\gamma^3).
\eea
Again, equations of motion are satisfied. 

Going to orders $\mathcal{O}(\gamma^2)$ and higher would require introduction of the most general terms preserving the $SO(n)$ symmetry both in the metric and the $B$ field as well as analysis of the resulting system of partial differential equations. We will not pursue this analysis here.

\section{Discussion}

In this article we have addressed two questions: extension of the group procedure for dualization of $S^3$ to other spheres and generalization of the well--known TsT transformations to the  non--abelian case. For both problems we constructed unique gravity solutions with the desired properties, although some of these geometries are not fully explicit. 

\bigskip

Let us briefly summarize our results. In section \ref{SecNATD} we proposed a gravitational counterpart of the non--abelian T duality for $n$--dimensional spheres, demonstrated the uniqueness of the proposed solution, and built them as expansions in powers of curvature. By construction, these solutions reduce to the abelian T dualities along $T^n$ when the radii of the spheres become large. The inspiration for this construction came from the three--dimensional sphere, for which the dual had been constructed in the past using worldsheet methods. Specifically, using the $S^3$ case as a guide, we imposed the invariance under $SO(n)$ in the ansatz (\ref{NatdSn}), then absence of singularities led to the unique ``gravitational T duals'' constructed in section \ref{SecSubNATDnew} for all values of $n$. 

We also proposed the gravitational counterparts of non--abelian TsT deformations for geometries containing $S^n\times S^n$ and $S^n\times T^n$ factors. Specifically, motivated by the known abelian TsT procedure for $T^n\times T^n$, we required the deformation to preserve the diagonal $SO(n)$ 
rotations acting on both elements of the products and to have only the NS--NS fields\footnote{In this article we focused on deformations of pure metrics. Extension of the ``non--abelian TsT deformations'' to geometries with Ramond--Ramond fluxes is an interesting open problem.}. 
The most general solutions of the relevant supergravity equations were found in section \ref{SecTsT}, and our main results were given by (\ref{GBinvS3S3}) for $S^3\times S^3$, by
 (\ref{SigmaGenYtSt}) and (\ref{TsTexpand}) for $S^3\times T^3$, by (\ref{AnswS4T4}) and (\ref{SnTnTsT}) for $S^n\times T^n$, and by (\ref{S4S4metr})--(\ref{S4S4bFrame}) 
 for $S^n\times S^n$ with $n\ne 3$. Although we mostly  focused on a one--parameter deformation which extends the TsT from $T^n\times T^n$ to product of spheres, in the latter case one can construct a larger class of interesting solutions, which have their origin in the torus case as well. Specifically, there are two $SO(n)$ invariant operations for $T^n\times T^n$: one can either perform a TsT transformation or just mix the coordinates of the tori. While formally this gives a two--parameter family of deformations, the mixture is usually discarded since it can be removed by a diffeomorphism. The situation with $S^n\times T^n$ and $S^n\times S^n$ is drastically different: the counterpart of the mixture of coordinates is no longer removable by coordinate transformations, so one has a genuine two--parameter family of deformations. An example of such family is analyzed in the Appendix \ref{AppS3T3extra}.

\bigskip

There are several natural extensions of our work. First, it would be interesting to extend the two--parameter deformation constructed in the Appendix \ref{AppS3T3extra} for 
$S^3\times T^3$ to products of spheres. The number of deformation parameters is expected to be the same for 
$S^n\times S^n$ and $S^n\times T^n$, and the second parameter reduces to a diffeomorphism only when both spheres are replaced by tori. Second, it would be interesting to find a non--abelian counterpart of the entire $O(2n,2n)$ duality group from the $T^n\times T^n$ reduction. In this article we focused on the $O(2,2)$ sector that preserved the diagonal $SO(n)$ rotations. This allowed us to reduce the problem to ODEs for $S^n\times T^n$ and to PDEs depending on two variables for $S^n\times S^n$. A counterpart of a generic $O(2n,2n)$ transformation would introduce nontrivial dependences on $n$ coordinates for $S^n\times T^n$ and $(2n)$ coordinates  for $S^n\times S^n$. While explicit solutions are probably out of reach, it would be interesting to understand their features at least qualitatively by exploring the relationship between the non--abelian TsT transformations and the formalism of Double Field Theory, which has been instrumental in understanding the general structure of the abelian case. Finally, the most interesting extension of our work would involve incorporation of the Ramond--Ramond fluxes and application of the non--abelian TsT transformations to RR backgrounds relevant for the gauge/gravity duality.

\section*{Acknowledgements}

This work was supported in part by the DOE grant DE-SC0015535.

\appendix

\section{Einstein's equations for the gravitational NATD}
\label{AppEinstNATD}
In this appendix we will provide some technical details of calculations associated with gravitational NATD along an $n$--dimensional sphere. The conceptual summary of the results is presented in section \ref{SecSubNATDnew}.

\bigskip

Let us consider the the geometry (\ref{NatdSn}) with  function $A$ given by (\ref{SnAfact}):
\bea\label{NatdSnApp}
ds^2&=&e^\psi\left[ds_{d-n-2,1}^2+dU^2+\frac{1}{U}\left(f_1 dr^2+f_2 r^2 d\Omega_{n-1}^2\right)\right]\,,\nn
e^{2\phi}&=&e^{-\frac{d-2}{2}\psi},\quad C^{(n-1)}=\frac{\sqrt{\beta} r^n f_4}{U^{\frac{n+1}{2}}}  d\Omega_{n-1};
\quad e^\psi=U^p f_3,\quad p=\frac{2n}{d-2}\,.
\eea
Furthermore, we will assume that functions $f_k$ depend on coordinates through a single combination $x$ given by (\ref{SnAfact}):
\bea
x=\frac{\eps^2 r^2}{U^2}
\eea
To analyze the Einstein's equations (\ref{EomNATD}) for the geometry (\ref{NatdSnApp}), it is convenient to split the Ricci tensor into three pieces as
\bea
R_{\mu\nu}=
R^{(1)}_{\mu\nu}+R^{(2)}_{\mu\nu}+w {\bar g}_{\mu\nu}\,,
\eea
where $R^{(1)}_{\mu\nu}$ has no derivatives of $\psi$, and neither $R^{(1)}_{\mu\nu}$ not $R^{(2)}_{\mu\nu}$ have components along the directions contained in $ds_{1,n-1}^2$. To isolate various terms and to assist with taking the decompactification limit later on, we temporarily make two modifications of the ansatz (\ref{NatdSn}):
\begin{itemize}
\item replace $e^\psi$ by $e^{\gamma\psi}$ in the metric
\item write $U=\frac{1}{\eps}+u$ as in (\ref{SnSeps})
\end{itemize}
This introduces two useful control parameters $(\gamma,\eps)$. 
Then direct calculations give
\bea
R^{(1)}_{uu}&=&-\frac{2 \eps^2 \left(-2 x^2  (f_1')^2+x  f_1 \left(4 x  f_1''+7  f_1'\right)+ f_1^2\right)}{({\eps} u+1)^2  f_1^2}\nn
&&-\frac{2 {\eps}^2 (n-1) \left(-2 x^2  (f_2')^2+x  f_2 \left(4 x  f_2''+7  f_2'\right)+ f_2^2\right)}{({\eps} u+1)^2  f_2^2}\,,\nn
R^{(1)}_{rr}&=&-\frac{2 {\eps}^2 \left(f_1^2-2 x^2  (f_1')^2+x  f_1 \left(4 x  f_1''+7  f_1'\right)\right)}{({\eps} u+1)^4  f_1}\nn
&&-\frac{(n-1) {\eps}^2 \left[x \left((4 x  f_1-1)  f_1'  f_2'+2  f_1  f_2''\right)+(2 x  f_1-1)  f_2 f_1'\right]}{({\eps} u+1)^4  f_1  f_2}\\
&&-\frac{(n-1){\eps}^2 \left( f_1  f_2 \left(2 x  f_2'+ f_2\right)+ f_2' \left(3  f_2-x  f_2'\right)\right)}{({\eps} u+1)^4  f_2^2}\,,\nn
R^{(1)}_{ru}&=&-\frac{2 {\eps}^2 (n-1) \sqrt{x} \left[ f_2 \left(x  f_1'  f_2'- f_1 
\left(2 x  f_2''+3  f_2'\right)\right)+ f_2^2  f_1'+x  f_1  (f_2')^2\right]}{({\eps} u+1)^3  f_1  f_2^2}\,,\nn
R^{(1)}_{ab}&=&\left\{\frac{x f_1'  \left(x f_2' +f_2 \right)-f_1  \left[\left(4 x^2 f_1' +5\right) xf_2' +2 x^2 f_2''+f_2  \left(2 x^2 f_1' +1\right)\right]}{f_1^2}\right.\nn
&&+\,1-4xf_2-8 x^3 f_2''-20 x^2 f_2'\nn
&&\left.+\frac{(3-n) \left(f_1  \left(4 x^3 (f_2')^2+f_2  \left(4 x^2 f_2' -1\right)+x f_2^2\right)+\left(x f_2' +f_2 \right)^2\right)}{f_1  f_2 }\right\}h_{ab}\,.\nonumber
\eea
Here and below prime denotes the derivative with respect to $x$. 
The contributions involving derivatives of the dilaton are
\bea
R^{(2)}_{uu}&=&-\frac{2 {\eps}^2 \gamma \left[20 n \gamma x  f_3'+ 
f_3 (4 \gamma \tau-5 n)\right]}{5 ({\eps} u+1)^2  f_3}\nn
&&-\frac{2 (d-2) {\eps}^2 \gamma \left[-30 (\gamma+2) x^2  (f_3')^2+15 x  f_3 \left(4 x  f_3''+5  f_3'\right)+16 \gamma f_3^2\right]}{15 ({\eps} u+1)^2  f_3^2}\,,
\nn
R^{(2)}_{rr}&=&\frac{2  {\eps}^2 n  {\gamma} \left(2 x  f_1'+ f_1\right)}{( {\eps} u+1)^4}
+\frac{(d-2)  {\eps}^2  {\gamma}x \left(f_1'-2f_1^2\right)f_3'}{( {\eps} u+1)^4  f_1  f_3}
\nn
&&+\frac{(d-2)  {\eps}^2  {\gamma} \left[( {\gamma}+2) x  (f_3')^2- f_3 \left(\left(4 x^2  f_1'+1\right)  f_3'+2 x  f_3''\right)\right]}{( {\eps} u+1)^4  f_3^2}\,,\\
R^{(2)}_{ru}&=&\frac{2  {\eps}^2 n  {\gamma}^2 \sqrt{x}  f_3'}{( {\eps} u+1)^3  f_3}
\nn
&&+\frac{(d-2)  {\eps}^2  {\gamma} \sqrt{x} \left( f_1 \left( f_3 \left(4 x  f_3''+3  f_3'\right)-
2 ( {\gamma}+2) x  (f_3')^2\right)-2 x  f_3  f_1'  f_3'\right)}{( {\eps} u+1)^3  f_1  f_3^2}\,,\nn
R^{(2)}_{ab}&=&\gamma\left\{ 2 n x \left[2 x  f_2'+ f_2\right]-\frac{(d-2) x  f_3' \left[x (4 x  f_1+1)  f_2'+(2 x  f_1+1)  f_2\right]}{ f_1 f_3}\right\}h_{ab}\,.\nonumber
\eea
Here 
\bea
\tau=-\frac{5n^2}{2(d-2)}-\frac{4}{3}(d-2).
\eea
Finally, the trace part is given by 
\bea
w&=&-\frac{2 \eps^2 n {\gamma} \left[f_1f_2 (2 n {\gamma}-n-1)-2 (n-1) x f_1f_2'-2 x f_2 f_1'\right]}{(d-2) (\eps u+1)^2 f_1 f_2}\nn
&&+\frac{2\eps^2 {\gamma}x (4 x f_1+1) [(f_3')^2-f_3f_3'']}{(\eps u+1)^2 f_1 f_3^2}+
\frac{\eps^2 {\gamma} f_3' [2 x f_1 (4 n {\gamma}-n-5)-n]}{(\eps u+1)^2 f_1 f_3}\nn
&&+\frac{\eps^2 {\gamma} x f_3' \left[(1-4 x f_1) f_2 f_1'-(n-1) (1+4 x f_1) f_1 f_2'\right]}{(\eps u+1)^2 f_1^2 f_2 f_3}\\
&&-\frac{(d-2) \eps^2 {\gamma}^2 x (4 x f_1+1) (f_3')^2}{(\eps u+1)^2 f_1f_3^2}\,.
\nonumber
\eea
To substitute these expressions into the Einstein's equations, one should set $\gamma=1$.

The Einstein's equations are invariant under rescaling of function $f_3$, which corresponds to a 
constant shift of the dilaton, so the order of derivatives in the dynamical equations can be lowered by introduction of a new  function $g_3$ through
\bea\label{f3g3App}
f_3=\exp\left[\int \frac{g_3 dx}{d-2}\right]\,.
\eea
Taking convenient linear combinations of the Einstein's equations, 
we observe that parameter $d$ disappears from the resulting system once it is rewritten in terms of $(f_1,f_2,f_4,g_3)$. In particular, for $n=3$, the final system gives the Einstein's equations in the string frame, so the absence of the $d$--dependence is expected. Interestingly, this feature persists for all values of $n$. 

The explicit form of the final equations for $(f_1,f_2,f_4,g_3)$ is not very illuminating, so we will just specify their main features. 
\begin{enumerate}[(a)]
\item There are five equations in total, and once they are satisfied, the equations of motion for the gauge field and scalar $\psi$ work as well, so no additional constraints are encountered.
\item Derivatives of function $f_4$ are present only in one equation, which is relatively simple:
\bea
&&f_2^{n-1} \left[2 f_1 (2 x g_3+n-1)+g_3+4(3n-1)xf_1'\right]
\nn
&&\qquad=\beta (n-1) \left[(2 (n+1) x f_1+n)f_4^2+2 x (4 x f_1+1) f_4f_4'\right]
\eea
\item Two equations contain $(f_1',f_2',f_1'',f_2'')$ and no other derivatives.
\item The last two equations contain $(f_1',f_2',g_3')$ and no other derivatives.
\end{enumerate}
Expanding all functions in power series in $x$ and substituting the results into the Einstein's equations, one finds 
{\it unique} solutions order--by--order in $x$. The first few terms are
\bea\label{PertArbN}
f_1&=&1+x \left[\frac{9 (n-1)}{7 (n+2)}+\frac{{\beta} \left(-5 n^2+10 n-3\right)}{14 (n+2)}\right]+\mathcal{O}(x^2)\nn
f_2&=&1+x \left[\frac{10 n+11}{7(n+2)}+\frac{{\beta} \left(-11 n^3-9 n^2+5 n+1\right)}{14 \left(n^2+n-2\right)}\right]+\mathcal{O}(x^2)\\
f_4&=&\beta+\beta x \left[\frac{19 n^2-19 n-14}{7 (n+2)}-\frac{{\beta} n \left(25 n^2-22 n+1\right)}{28 (n+2)}\right]+\mathcal{O}(x^2)\nonumber
\eea
Subsequent terms can be computed to arbitrary orders, but unfortunately there are no recognizable patterns for $n\ne 3$. In the case of the three--dimensional sphere, we find
\bea\label{f1234ExpS3}
f_1&=&1-\frac{9{\bar\beta}x}{35}+\frac{3{\bar\beta}(3920+3307{\bar\beta})x^2}{53900}+\mathcal{O}(x^3)\nn
\frac{1}{f_2}&=&1+4x+\frac{181{\bar\beta}x}{70}-\frac{{\bar\beta}(67270+32313{\bar\beta})x^2}{13475}+
\mathcal{O}(x^3)\\
\frac{1}{f_3}&=&1+4x+3{\bar\beta}x-\frac{3{\bar\beta}(42+17{\bar\beta})x^2}{35}+\mathcal{O}(x^3)\nn
\frac{\sqrt{\beta}}{f_4}&=&1+4x+\frac{24{\bar\beta}x}{7}-\frac{3{\bar\beta}(46760+5171{\bar\beta})x^2}{53900}+
\mathcal{O}(x^3),
\nonumber
\eea
with ${\bar\beta}=\beta-2$, 
so the NATD (\ref{NatdS3}) is recovered for $\beta={2}$.

Numerical integration discussed in section \ref{SecSubNATDnew} demonstrates that the geometry (\ref{NatdSn}) encounters a naked curvature singularity at some finite value of $r$, which is located a finite distance away from $r=0$, unless parameter beta takes the values ${\hat\beta}_n$ presented in Table \ref{Table1}. For $\beta={\hat\beta}_n$, the regular geometry defined for all $r$ can be interpreted as a ``geometric T dual'' of the sphere $S^n$, a generalization of the well--known solution (\ref{NatdS3}) beyond $n=3$. To analyze the behavior of this geometry at infinity, it is useful to construct asymptotic expansions of various functions at large $r$:
\bea\label{InfExpandApp}
f_1&=&\frac{c_2 \left[9n-2+\zeta (n-1)\right]}{4 (n-2) \sqrt{x}}+
\frac{2n- (n-1)^2\zeta}{4 x}+\mathcal{O}(x^{-3/2})\,,\nn
f_2&=&\frac{c_2}{\sqrt{x}}-\frac{(n-2) \left[ (n-2) (n-1)\zeta-6 n\right]}{x 
\left[9n-2+\zeta (n-1)\right]}+\mathcal{O}(x^{-3/2}),\\
g_3&=&\frac{1}{x}\left[n-\frac{(n-2) \left[\zeta (n-1)^3-2 n(3n-2)\right]}{2 c_2 \sqrt{x} 
\left[9n-2+\zeta (n-1)\right]}+\mathcal{O}(x^{-1})\right]\,,\nn
f_4&=&x^\frac{n+1}{2}\left[c_4+\frac{c_4 (n-1) (5n-2n^2-2)}{2 c_2 \sqrt{x} 
\left[9n-2+\zeta (n-1)\right]}+\mathcal{O}(x^{-1})\right]\nonumber
\eea
Here we defined a convenient parameter
\bea
\zeta=c_4^2c_2^{1-n}\,.
\eea
Expansions (\ref{InfExpandApp}) can be continued to an arbitrary order in $x^{-1}$, and they are fully specified by three parameters $(\beta,c_2,c_4)$. The first parameter takes the critical value ${\hat\beta}$, while the values of $(c_2,c_4)$ are uniquely determined by numerical integration of equations starting with expansions (\ref{PertArbN}) near $x=0$. Keeping only the 
leading terms in the expansions (\ref{InfExpandApp}), one arrives at a very simple asymptotic geometry (\ref{S4asympInf}).

\section{$SO(n)$--invariant deformations of $T^n\times T^n$}
\label{AppTor}

Although the main goal of this article is to explore deformation of geometries involving spheres, 
$S^n\times S^n$, to gain some intuition it is also useful to study the degenerate limits where the warp factor in front of at least one of the spheres goes to infinity. One such limit gives $S^n\times T^n$ analyzed in sections 
\ref{SecSubS3T3}--\ref{SecSubSnTn}, and in the second limit one gets $T^n\times T^n$. In this appendix we will look at this simple setting and construct the most general deformation of such space. This analysis culminating in explicit solutions provides some insights into deformations of $S^n\times T^n$ and $S^n\times S^n$ explored in section \ref{SecTsT}.

\bigskip

We begin with looking at a decompactification limit of the metric (\ref{SigmaGenY}) which is obtained by replacing the left--invariant forms $\sigma_a$ by three flat directions $x_a$:
\bea\label{SigmaGenYtor}
ds^2=f_1 dx_a dx_a+\frac{1}{f_2}(dy_a+f_4 dx_a)(dy_a+f_4 dx_a)+
ds_\perp^2,
\quad B=0,\quad e^{2\Phi}=1.
\eea
By construction, there is no dependence on $(x_a,y_a)$ directions. Although index $a$ here runs from one to three, it is easy to see that the same metric emerges in the decompactification limit (\ref{TnSnDef}) in higher 
dimensions\footnote{Recall that in the decompactification limit one writes $x_a=\frac{1}{\eps}+{\bar x}_a$ and sends parameter $\eps$ to infinity of the geometry after rescaling various functions of transverse coordinates.}, so we will analyze the metric (\ref{SigmaGenYtor}) with index $a$ running from $1$ to $n$. As in sections \ref{SecSubS3T3}--\ref{SecSubSnTn} we will focus on flat transverse space $ds_\perp^2$ and assume that $(f_1,f_2,f_4)$ are functions of only one coordinate $r$. Then performing series expansions of these functions around some point $r=r_0$, one finds that 
function $f_4$ can be removed by a linear transformation of $(x,y)$ with {\it constant coefficients}, i.e., without breaking translational invariance in $(x_a,y_a)$. Therefore, after an appropriate redefinition of coordinates, one arrives at the diagonal form of the metric (\ref{SigmaGenYtor}):
\bea\label{tempAug20}
ds^2=dr^2+{F}^2(dx_a)^2+{G}^2 (dy_a)^2\,.
\eea
Invariance under simultaneous rescalings of functions $(F,G)$ and coordinates of the tori suggest a change of variables which lowers the order of the Einstein's equations:
\bea
F=\exp\left[\int A dr\right],\quad G=\exp\left[\int B dr\right].
\eea
The equations themselves are 
\bea
A^2+\frac{2n}{n-1} AB+B^2=0,\quad
{\dot A}=-nA(A+B),\quad {\dot B}=-nB(A+B)
\eea
and the most general solution reads
\bea
A=\frac{1+\mu}{2n(r+r_0)},\quad B=\frac{1-\mu}{2n(r+r_0)}\quad \mu=\pm \sqrt{2n-1}\,,
\eea
where $r_0$ is an integration constant. Adjusting two additional constants appearing in the integrals for $F$ and $G$ by rescaling the tori coordinates, we arrive at the most general solution of the Einstein's equations for the metric (\ref{tempAug20}):
\bea\label{tempAug20soln}
F=\left[1+\frac{r}{r_0}\right]^{\frac{1+\mu}{2n}},\quad
G=\left[1+\frac{r}{r_0}\right]^{\frac{1-\mu}{2n}},\quad \mu=\pm \sqrt{2n-1}\,.
\eea
More generally, for the metric of the form
\bea
ds^2=dr^2+\sum_{k=1}^p (F_k)^2 (dx_k)^2
\eea
where all $F_k$ are functions of one radial coordinate, we can introduce new variables $A_k$ by
\bea
F_k=\exp\left[\int A_k dr\right].
\eea
Then the Einstein's equations reduce to 
\bea\label{tampAug20a}
{\dot A}_k=-A_k \sum_j A_j
\eea
and one additional constraint which will be specified below. Adding all equations (\ref{tampAug20a}), we arrive at a decoupled relation for $[\sum A_j]$, which has the most general solution
\bea\label{tampAug20b}
\sum A_j=\frac{1}{p}\frac{1}{(r+r_0)}\,.
\eea
Then substitution into (\ref{tampAug20a}) gives
\bea
A_k=\frac{a_k}{2(r+r_0)}\,.
\eea
with undetermined constants $(a_k,r_0)$. Substituting this into the remaining Einstein's equations, we arrive at a set of algebraic relations for parameters $a_k$:
\bea
\sum a_k=2,\quad \sum (a_k)^2=4,
\eea
and at the most general solution of Einstein's equations:
\bea
F_k=\left[1+\frac{r}{r_0}\right]^{a_k/2}.
\eea
In the case of two blocks (\ref{tempAug20}) we recover (\ref{tempAug20soln}).

\bigskip
 
Let us now explore the relation between the explicit solution (\ref{tempAug20soln}) and deformations of 
$S^n\times T^n$ constructed in section \ref{SecSubSnTn}. To avoid unnecessary complications, we focus on $n=3$, and other spheres work in the same way. 

The Einstein's equations for the metric (\ref{SigmaGenYnew}) give (\ref{EinstS3T3simple}), as well as one more equation
\bea\label{EinstS3T3compl}
&&2r^2[1-(rh)^4]{\ddot h}+3 g^2 h^7 r^6-h^5 r^4 \left[2 g^2+4 g \left({\dot h}^2 r^4+7\right)-3 {\dot h}^2 r^4+8\right]\nn
&&\quad-2 h^4 {\dot h} r^5 \left[-2 g^2-10 g+{\dot h}^2 r^4+1\right]-h^3 r^2 \left[17 g^2+48 g-4 {\dot h}^2 r^4+16\right]\\
&&\quad-h \left[2 g^2-4 g+{\dot h}^2 r^4\right]+2 \left(2 g^2+6 g+5\right) {\dot h} r-2 (g-3) g h^6 {\dot h} r^7-2 g h^2 {\dot h} r^3=0.\nonumber
\eea
In the decompactification limit of $S^3$, the metric becomes diagonal, then equation (\ref{EinstS3T3compl}) is trivially satisfied, and the most general solution of the system (\ref{EinstS3T3simple}) is given by (\ref{Spurious}). Note that this solution is obtained in the gauge 
\bea
ds^2=f d{\bar r}^2+{{\bar r}^2} (dx_a dx_a)+G^2 dy_a dy_a, 
\eea
which is different from (\ref{tempAug20}). Recall that 
\bea
G=\exp\left[\int \frac{g}{\bar r} d{\bar r}\right],\quad f=1+3g+g^2\,.
\eea
The map to (\ref{tempAug20}) is given by
\bea
F={\bar r},\quad dr=\sqrt{f}d{\bar r}\,.
\eea
To recover the answer (\ref{tempAug20soln}), we expand the solution (\ref{Spurious}) for large values of ${\bar r}$:\footnote{Recall that the decompactification limit (\ref{DecompT4S4}) is obtained by scaling the radial coordinate as $\eps^{-1}$ while keeping all constants, such as ${\bar r}_0$, fixed.}
\bea
\frac{\bar r}{{\bar r}_0}=-\frac{3-\sqrt{5}}{4\sqrt{5}}\ln\left[(3-\sqrt{5})\,z\right]+
\frac{3+\sqrt{5}}{4\sqrt{5}}\ln\left[3\sqrt{5}-5\right]-\frac{3(2-\sqrt{5})}{10}z+\mathcal{O}(z^2)\nonumber
\eea
Here we defined 
\bea
z=-g+g_0,\quad g_0=-\frac{3-\sqrt{5}}{2}
\eea
In particular, in this approximation,
\bea
G\simeq {\bar r}^{g_0}=F^{-\frac{3-\sqrt{5}}{2}}\,,
\eea
in the perfect agreement with (\ref{tempAug20soln}) for $n=3$. The expected relation between $F$ and $r$ is recovered as well.

\section{Two--parameter deformation of $S^3 \times T^3$}
\label{AppS3T3extra}

In section \ref{SecSubS3T3} we constructed a one--parameter deformation of the $S^3 \times T^3$ geometry. Depending on the duality frame, the answer can be viewed either as a pure metric or as a result of the ``non--abelian TsT transformation''. In contrast to the abelian case, even the metric deformation is highly nontrivial since it does not reduce to a simple shift. Therefore, it is interesting to combine the metric deformation and the ``non--abelian TsT transformation'' into a more general solution with two deformation parameters. To do that, we consider the following ansatz in the string frame
\bea\label{S3S3TwoPar1}
ds^2&=&dr^2+\frac{(r f_1)^2}{4}\bigg[\sigma_a+\beta f_5 dy_a\bigg]^2
+\frac{(f_2)^2}{4}dy_a dy_a,\nn
B&=&\gamma f_4 \sigma_a\wedge dy_a,\quad e^{2\phi}=f_3.
\eea
Solving the Einstein's equations, we arrive at the unique expansions: 
\bea\label{S3S3TwoPar1a}
f_1&=&1 - 8 (\gamma r)^2  - 
 \frac{r^4}{30} [1472 \gamma^4 + 448 (\beta\gamma)^2 - 3 \beta^4]+\mathcal{O}(\alpha^6),\nn
 f_2&=&1 - 8 (\gamma r)^2  - 
 \frac{r^4}{6}[192 \gamma^4 + 128(\beta\gamma)^2 + \beta^4]+\mathcal{O}(\alpha^6),\nn
 f_3&=&1 - (4\gamma r)^2 -32(\beta\gamma)^2 r^4+\mathcal{O}(\alpha^6),\\
 f_4&=&r^2+ \frac{4r^4}{3}[\beta^2-4 \gamma^2]+ 
 \frac{r^6}{30} [256 \gamma^4 + 224 (\beta\gamma)^2 + 51 \beta^4]+\mathcal{O}(\alpha^6),\nn
 f_5&=&1 + \frac{2r^2}{3} [32 \gamma^2 + \beta^2] + 
 \frac{r^4}{6}[2816 \gamma^4 + 352 (\beta\gamma)^2 + 3 \beta^4]+\mathcal{O}(\alpha^6).
 \nonumber
\eea
Here $\alpha=\sqrt{\beta^2+\gamma^2}$ is a parameter that controls the strength of the deformation.  In the decompactification limit,
\bea
r=\frac{1}{\eps}+{\bar r},\quad \sigma_a\rightarrow \eps dx_a,\quad \beta\rightarrow \eps{\bar \beta},\quad \gamma\rightarrow \eps{\bar \gamma}\,,
\eea
we find a combination of an abelian TsT transformation and and a shift:
\bea
ds^2&=&dr^2+\frac{(1+\eps {\bar r})^2 ({\bar f}_1)^2}{4}\bigg[dx_a+{\bar\beta} {\bar f}_5 dy_a\bigg]^2
+\frac{({\bar f}_2)^2}{4}dy_a dy_a+\mathcal{O}(\eps^2),\nn
B&=&{\bar\gamma} {\bar f}_4 \sigma_a\wedge dy_a,\quad e^{2\phi}={\bar f}_3.
\eea
To make the symmetry under T duality more explicit, we can define ${\tilde\beta}=\beta/4$ and redistribute the complete squares in (\ref{S3S3TwoPar1}) to rewrite the solution in the form
\bea\label{S3S3TwoPar2}
ds^2&=&dr^2+\frac{(r h_1)^2}{4}(\sigma_a)^2
+\frac{(h_2)^2}{4}[dy_a +4{\tilde\beta} h_5 \sigma_a]^2,\nn
B&=&\gamma h_4 \sigma_a\wedge dy_a,\quad e^{2\phi}=h_2 h_3.
\eea
The functions appearing in this expression are given by
\bea\label{S3S3TwoPar2a}
h_1&=&1-{8r^2}[{\tilde\beta}^2+\gamma^2]-\frac{32r^4}{15}
\left[23{\tilde\beta}^4+242({\tilde\beta}\gamma)^2+23\gamma^4\right]+\mathcal{O}(\alpha^6),\nn
h_2&=&1+{8r^2}[{\tilde\beta}^2-\gamma^2]+{32r^4}
\left[3{\tilde\beta}^4-2({\tilde\beta}\gamma)^2-\gamma^4\right]+\mathcal{O}(\alpha^6),\nn
h_3&=&1-{8r^2}[{\tilde\beta}^2+\gamma^2]-{32r^4}
\left[{\tilde\beta}^4+14({\tilde\beta}\gamma)^2+\gamma^4\right]+\mathcal{O}(\alpha^6),\\
h_4&=&r^2+ \frac{16r^4}{3}[4{\tilde\beta}^2-\gamma^2]+ 
 \frac{128r^6}{15} [\gamma^4 + 14 ({\tilde\beta}\gamma)^2 + 51 {\tilde\beta}^4]+\mathcal{O}(\alpha^6),\nn
h_5&=&r^2+ \frac{16r^4}{3}[4\gamma^2-{\tilde\beta}^2]+ 
 \frac{128r^6}{15} [51\gamma^4 + 14 ({\tilde\beta}\gamma)^2 + {\tilde\beta}^4]+\mathcal{O}(\alpha^6)\,.\nonumber
\eea
The T duality along the 3--torus acts on these ingredients in the expected way:
\bea
T:\quad {\tilde\beta}\leftrightarrow\gamma,\quad (h_1,h_2,h_3,h_4,h_5)\rightarrow (h_1,\frac{1}{h_2},h_3,h_5,h_4).
\eea
Although in this appendix we focused on deformations of $S^3 \times T^3$, solutions similar to (\ref{S3S3TwoPar2})--(\ref{S3S3TwoPar2a}) can be constructed for all $S^n \times T^n$ by combining two deformations (\ref{AnswS4T4}) and (\ref{SnTnTsT}) into a larger family.

\section{Deformations of $S^2\times T^2$}
\label{AppS2T2}

In this short appendix, we will construct the deformation of $S^2\times T^2$ in the first two orders in $\gamma$. In contrast to the generic case discussed in section \ref{SecSubSnTn}, now the metric can contain the Levi-Civita symbol, and this leads to some technical complications, although conceptually the analysis is very similar to the one presented in section \ref{SecSubSnTn}. We will see that {\it once the Einstein's equations are solved}, the additional terms with 
$\eps_{ab}$ can be removed by rotating the torus coordinates without touching those of the sphere. Therefore, in spite of somewhat different intermediate steps, the final result for $S^2\times T^2$ is identical to that for $S^n\times T^n$ with $n>3$, and both these cases are distinct from the deformations of $S^3\times T^3$ studied in section \ref{SecSubS3T3}. 

\bigskip

As discussed in section \ref{SecSubSnTn}, the metric of the deformed $S^n\times T^n$ must be built from the ingredients (\ref{Aug15}), and the Levi-Civita symbol can appears at most once in every term. For $n>3$, this leads to the most general ansatz (\ref{TnSnDef}), but for $S^2\times T^2$ one encounters several additional terms. In this case, the most general metric built from the ingredients (\ref{Aug15}) is
\bea\label{S2T2anstz}
ds^2&=&dr^2+r^2 f_1(dx_a)^2+r^2 f_2(x_a dx_a)^2+f_3 (dy_a)^2+f_4(x_a dy_a)^2\nn
&&+\gamma^2 r^2 f_5 (x_adx_a)(\eps_{bc}x_b dx_c)+\gamma^2 f_6 (x_ady_a)(\eps_{bc}x_b dy_c)\nn
&&+\gamma r^2\left[h_1\delta_{ab}+
h_2 x_a x_b+h_3 x_a\eps_{bc}x_c+h_4 x_b\eps_{ac}x_c+h_5\eps_{ab}\right]
dx_ady_b
\eea
Here $(f_1,\dots f_6)$ and $(h_1,\dots h_5)$ are functions of $r$ and $\rho=\sqrt{(x_a)^2}$. The terms with $(f_5,f_6,h_3,h_4,h_5)$ appear only in the $n=2$ case discussed here. 
Note that due to the identity
\bea
x_a\eps_{bc}x_c-x_b\eps_{ac}x_c-\rho^2 \eps_{ab}=0,
\eea
one combination of functions $(h_3,h_4,h_5)$ is redundant. Therefore, we can set $h_3=0$.

A priori all functions in (\ref{S2T2anstz}) depend on $r$ and $\rho$, so in the first order in $\gamma$ one arrives at a system of equations for $h_k$ similar to (\ref{SnTnEinst}). However, as discussed in section \ref{SecSubSnTn}, it is natural to fix the $r$ dependence using the abelian TsT as a guide, and in the parameterization (\ref{S2T2anstz}) this means assuming that $(f_k,h_k)$ depend only on 
$\rho$. Then solving the Einstein's equations in the first order in $\gamma$ we find
\bea\label{S2T2ord1}
ds_\gamma^2=\frac{\gamma r^2}{[1+\rho^2]^2}\left[c_1(1-\rho^2)\delta_{ab}+
2c_1 x_a x_b+c_2(1+\rho^2)\eps_{ab}+c_2\eps_{ac}x_b x_c \right]dx_a dy_b
\eea
There are two more solutions which diverge at $\rho=0$ and thus don't have a good decompactification limit. We have set the corresponding integration constants to zero. 

In the decompactification limit (\ref{DecompLim}) the metric (\ref{S2T2ord1}) becomes
\bea\label{S2T2ord1shft}
ds^2=d{\bar r}^2+(1+\eps{\bar r})^2\left[2d{\bar x}_a+
\frac{{\bar\gamma} c_1}{2}dy_a+\frac{{\bar\gamma} c_2}{2}\eps_{ab}dy_b\right]^2+dy_bdy_b
+\mathcal{O}(\eps^2)+\mathcal{O}(\gamma^2)\,,
\eea
so the deformation (\ref{S2T2ord1}) describes the standard shift 
${\bar x}_a\rightarrow {\bar x}_a+\frac{{\bar\gamma} c_1}{2}y_a$ as well as the shift related by the two--dimensional Hodge duality, 
${\bar x}_a\rightarrow {\bar x}_a+\frac{{\bar\gamma} c_2}{2}\eps_{ab}y_b$. The latter option is a peculiarity of the two--dimensional case that has not been encountered in section \ref{SecSubSnTn}. The ``dual shift'' can be eliminated by an $SO(2)$ rotation,
\bea
y_a\rightarrow {\tilde y}_a=\frac{1}{\sqrt{c_1^2+c_2^2}}[c_1 y_a+{c_2}\eps_{ab}y_b]
\eea
so without loss of generality, we can set $(c_1,c_2)=(1,0)$ in (\ref{S2T2ord1shft}) and (\ref{S2T2ord1}). Then (\ref{S2T2ord1}) reduces to (\ref{TsTS4pert}), and the subsequent analysis proceeds as in section \ref{SecSubSnTn}. For example, solving the Einstein's equations in the second order in $\gamma$ we find 
\bea\label{S2T2final}
ds^2&=&dr^2+r^2\frac{4dx_adx_a}{[1+\rho^2]^2}+\left[1+\frac{(\gamma r)^2}{4}-
\frac{(c+1)(\gamma r\rho)^2}{2(1+\rho^2)^2}
\right]dy_bdy_b+\frac{c(\gamma r)^2}{(1+\rho^2)^2}(x_a dy_a)^2\nn
&&+\frac{2\gamma r^2}{(1+\rho^2)^2}\left[2(x_a dx_a)(x_b dy_b)
+(1-\rho^2)dx_ady_a\right]\\
&&+\frac{c_1\gamma^2r^4}{(\rho^2+1)^4}\left[12(x_a dx_a)^2+(\rho^4-10\rho^2+1)(dx_a)^2\right]\nn
&&-
\frac{3c_1\gamma^2r^2}{4(\rho^2+1)^2}\left[\rho^4-4\rho^2+1\right](dx_a)^2+\mathcal{O}(\gamma^3).\nonumber
\eea
For $c_1=0$, this recovers the solution (\ref{AnswS4T4}) with $\eta$ given by (\ref{SnTnEta}) for $n=2$. The last two lines 
in (\ref{S2T2final}) give an additional potential deformation at the quadratic order that is peculiar for $n=2$. For $c_1$, the solution (\ref{S2T2final}) can be extended to a deformation of $S^2\times S^2$ yielding either the TsT deformation (\ref{S4S4bFrame}) or its metric version (\ref{S4S4metr}). Similar extensions can be constructed even for arbitrary $c_1$, but we will not discuss them further.

\end{document}